\documentclass[lettersize,journal]{IEEEtran}
\usepackage{graphicx} 

\usepackage{xcolor}
\usepackage{amsthm}
\usepackage{amsmath,amsfonts}
\usepackage{amssymb}

\usepackage{array}
\usepackage{textcomp}
\usepackage{stfloats}
\usepackage{url}
\usepackage{verbatim}
\usepackage{cite} 
\usepackage[english]{babel}

\theoremstyle{definition}

\usepackage{mathtools}
\usepackage{icomma}
\usepackage{verbatim}
\usepackage{caption}
\usepackage{multirow}
\usepackage{booktabs}  
\usepackage{siunitx}
\usepackage{graphicx}
\usepackage{comment}
\usepackage{svg}
\usepackage{tabularx}
\newtheorem{proposition}{Proposition}
\usepackage[dvipsnames]{xcolor}

\usepackage{algorithm}
\usepackage[noend]{algpseudocode}

\theoremstyle{remark}

\usepackage{subcaption} 

\usepackage{hyperref}
\hypersetup{
    colorlinks=true,
    linkcolor=blue,
    filecolor=magenta,      
    urlcolor=cyan,
    pdftitle={Overleaf Example},
    pdfpagemode=FullScreen,
    }

\begin{document}

\title{Two-dimensional Decompositions of High-dimensional Configurations for Efficient Multi-vehicle Coordination at Intelligent Intersections}

\author{Amirreza Akbari, Johan Thunberg,~\IEEEmembership{Member,~IEEE}

\thanks{This paper is under review at IEEE Transactions on Intelligent Transportation Systems. Copyright may be transferred without notice, after which this version may no longer be accessible.}

\thanks{The authors are with the Department of Electrical and Information Technology (EIT), Lund University, SE-22100 Lund, Sweden (e-mail: amirreza.akbari@eit.lth.se; johan.thunberg@eit.lth.se).}
\thanks{This work was partially supported by the Wallenberg AI, Autonomous Systems and Software Program (WASP) funded by the Knut and Alice Wallenberg Foundation.}
}



\maketitle

\begin{abstract}
For multi-vehicle complex traffic scenarios in shared spaces such as intelligent intersections, safe coordination and trajectory planning is challenging due to computational complexity. To meet this challenge, we introduce a computationally efficient method for generating collision-free trajectories along predefined vehicle paths. We reformulate a constrained minimum-time trajectory planning problem as a problem in a high-dimensional configuration space, where conflict zones are modeled by high-dimensional polyhedra constructed from two-dimensional rectangles. Still, in such a formulation, as the number of vehicles involved increases, the computational complexity increases significantly. To address this, we propose two algorithms for near-optimal local optimization that significantly reduce the computational complexity by decomposing the high-dimensional problem into a sequence of 2D graph search problems. The resulting trajectories are then incorporated into a Nonlinear Model Predictive Control (NMPC) framework to ensure safe and smooth vehicle motion. We furthermore show in numerical evaluation that this approach significantly outperforms existing MILP-based time-scheduling; both in terms of objective-value and computational time. 
\end{abstract}


\section{Introduction}
\IEEEPARstart{I}{ntersections} is a complicated scenario in traffic management, accounting for approximately 20 percent of all road traffic fatalities in the European Union \cite{sander2018market} and 44 percent of all reported traffic incidents in the United States \cite{6843706}. Thus, intersection management plays a critical role in transportation systems \cite{olayode2020intelligent}. To improve safety and efficiency at road intersections, various approaches have been proposed in the Intelligent Transportation Systems (ITS) literature. Generally, intersections can be categorized into three types: signalized, autonomous, and hybrid \cite{9678327}. 

In autonomous and hybrid intersections, vehicles communicate with each other or with the infrastructure using Vehicle-to-Everything (V2X) technology \cite{7992934}. A major focus of research in this area is the design and improvement of scheduling policies for Intersection Management (IM) \cite{khayatian2020survey}. In addition to the First-Come-First-Served (FCFS) method, two main classes of scheduling approaches have been studied: analytical and heuristic algorithms. Heuristic methods offer fast, though not necessarily optimal, solutions, while analytical-based algorithms aim for optimality but could be more time-consuming.

\subsection{Related Works}

\subsubsection{Analytical Approaches}
There are numerous studies on analytical scheduling algorithms \cite{medina2019optimal, fayazi2018mixed, yao2020integrated, 10007069, ahn2017safety}. In \cite{10007069}, the authors propose a two-stage optimization method for managing Connected Autonomous Vehicles (CAVs) at an isolated intersection. The first stage uses a Mixed-Integer Linear Program (MILP) to generate a conflict-free vehicle arrival schedule, while the second stage applies linear programming (LP) to optimize eco-driving trajectories. This two-stage MILP–LP framework reduces vehicle delays and fuel consumption. In \cite{yao2020integrated}, the authors present a two-level optimization approach for CAVs in conflict zones. At the upper level, a binary MILP schedules vehicle entries into the zone, while the lower level refines these results using a nonlinear programming model, transformed via infinitesimal methods, to derive optimal trajectories. In \cite{fayazi2018mixed}, the intersection crossing scheduling problem is modeled as an MILP, where an intersection server periodically assigns optimal arrival times to CAVs. A separate trajectory planner adjusts each vehicle’s speed to meet the scheduled arrival time. In \cite{medina2019optimal}, the authors develop a hybrid queuing model and propose an MPC-based approach to optimize the vehicle crossing sequence. The low-level control is managed by a Cooperative Intersection Control (CIC) system, while the high-level MPC minimizes the total time spent in the intersection. In \cite{10976401}, the authors introduce OPC-CBS, which merges conflict-based search with optimal control. The method transforms a global optimization problem into smaller multi-stage problems, enabling real-time spatiotemporal conflict resolution with computational efficiency and energy savings. In \cite{10927034}, a hybrid analytic approach is employed using Petri nets to avoid conflicts and deadlocks, followed by iterative trajectory refinement. This structured model supports large-scale multi-lane intersection coordination with reduced computational cost.

\subsubsection{Heuristic Approaches}

For heuristic algorithms, the following works are notable. In \cite{wu2019dcl}, the authors introduce DCL-AIM, a decentralized coordination learning approach for autonomous intersection management. The vehicle coordination problem is modeled as a Multi-Agent Markov Decision Process (MAMDP) and solved via Multi-Agent reinforcement learning (MARL). The authors decompose the decision-making problem to exploit sparse agent interactions and use decentralized coordination. In \cite{mitrovic2019combined}, a hybrid framework is proposed that combines alternate-direction lane assignment with a reservation-based scheduling system for intersection control. Vehicles are dynamically assigned to lanes with alternating flow directions to maximize throughput, and a conflict-point-based reservation mechanism determines the crossing sequence using a backpressure-like policy. In \cite{10403872}, the authors formulate a constrained optimization problem to compute collision-free cooperative trajectories at unsignalized intersections. They develop a learning-based iterative optimization (LBIO) method to pretrain velocity profiles, followed by a Monte Carlo Tree Search (MCTS) algorithm that dynamically determines the optimal vehicle passing sequence. Authors in \cite{10774177} introduce MA-GA-DDPG, a multi-agent reinforcement learning method incorporating attention mechanisms and hierarchical game priors. The attention module identifies relevant interaction agents, while the game-theoretic layer models traffic priority. A safety inspector refines the learned policies, enabling safer and more efficient interactions among CAVs. \cite{10858594} presents a hierarchical framework using graph attention networks to predict surrounding traffic behavior and guide an MPC-based low-level planner. This learning-based approach focuses on mutual interactions in dense traffic scenarios.

\subsection{Contribution}

To enhance safety and efficiency at road intersections, this paper introduces a trajectory planner combined with a controller that works in sequence. Although some studies show that vehicles may not strictly follow fixed lanes inside intersections, leading to path variations \cite{10115056}, \cite{zhao2023microscopic}, the approach in this work assumes predefined paths for each vehicle. This assumption allows the planner to focus on coordinating the timing along these paths to ensure safe and efficient traversal of the intersection. The planning algorithm computes a near-optimal, collision-free trajectory by coordinating traversal of these pre-defined paths within the shared space. Rather than generating entirely new routes, the algorithm produces trajectories by focusing on adjusting the timing or sequencing along the given paths to avoid collisions and ensure safe navigation through potential conflict regions.

To compute such trajectories, the problem is formulated in a \textit{configuration space} \cite{lavalle1998optimal, latombe2012robot}, which encapsulates the configurations of the vehicles. In a system with $N$ vehicles, the configuration space becomes N-dimensional. Graph search to solve shortest path problems for visibility graphs in the high-dimensional configuration space could be used, but finding, representing and searching such graphs in a fast manner is computationally expensive; both time and memory requirements scale as $\mathcal{O}(k^N)$, where $k$ is the number of discretization steps per trajectory~\cite{katrakazas2015real}. Moreover it is known that in general motion planning in polyhedra obstacle environments is NP-hard \cite{canny1988complexity}.

To address this, the proposed methodology avoids explicitly constructing the full joint multi-vehicle configuration space. Instead, it leverages a \textit{dimensionality decomposition} technique. Specifically, it decomposes the N-dimensional problem into a sequence of $(N - 1)$ two-dimensional (2D) problems, where each corresponds to a combination of a subset of vehicles. This is based on the observation that in 2D spaces with rectangular obstacles, optimal paths either pass obstacle (rectangle) corners or boundaries or reside in the free space. Furthermore, vehicles are assumed to move forward (or stand still) along their paths, which has the implication that each visibility graph in each 2D subspace of the $N$-dimensional configuration space is a directed and acyclic graph (DAG). 

Thus, the computational complexity (time) for solving the shortest path problem is $\mathcal{O}( r^2)$ where $r$ is the number of conflict regions. For each conflict region there is a rectangle that contributes with at most four nodes (i.e., corners of the rectangle) to the visibility graph. Thus, the computational time is faster than that of for example Dijkstra's algorithm. As shown in~\cite{lavalle2006planning}, a generalization from 2D to higher dimension becomes increases computational complexity significantly, which emphasizes the benefit of operating in 2D subspaces. In our approach, there are $N - 1$ 2D-problems and the total computational complexity of a single coordination sequence (trajectory planning problem) is $\mathcal{O}(N r^2)$, where $r$, again, is the number of conflict regions.

However, the quality of the solution (the generated path in the $N$-dimensional configuration space), depends on the order of the sequence of the $(N-1)$ 2D-problems or selected trajectories, i.e.,  the order of the selection of subsets of vehicles. Such an order can be represented by a permutation, and there are $N!$ many such.  There are different ways of creating subsets of vehicles or equivalently checking permutations of the order. We investigate two such, where the number permutations in the fist is $\frac{N!}{2}$ and the number of permutations in the second approach is $\frac{N!}{\left\lvert \operatorname{Aut}(T_N) \right\rvert}$, where $\operatorname{Aut}(T_N)$ denotes the automorphism group of the coordination tree $T_N$. 
The coordination tree $T_N$ is a hierarchical structure that represents how the trajectories of $N$ vehicles are grouped and coordinated step by step.

The computed path in the $(N-1)$-dimensional configuration space is used to computed trajectories. 
The path is in configuration in $(N-1)$-dimensional configuration space is represented by a scaled time-varying unit vector, where the $i$'th element thereof corresponds to the velocity of vehicle $i$ along its path. 

Subsequently, the elements of the vector are used for tracking in a completely decoupled nonlinear model predictive control (NMPC) formulation that produces dynamic controllers. Below we shortly list our contributions. 

    \begin{itemize}
        \item   
        We provide a computationally fast multi-vehicle minimum-time constrained trajectory planning procedure by casting the problem as a shortest path problem in an high-dimensional configuration space, which is subsequently addressed by decomposing the problem into a sequence of 2D graph search problems. We introduce two distinct decomposition algorithms and provide a comparative analysis of their performance. The procedure is shown, in numerical experiments, to outperform a MILP-based time-schedule-based approach. 
        \item     We demonstrate how the computed high-dimensional trajectories, represented by a a scaled time-varying unit vector, can be used in an NMPC framework that provides controllers for tracking the the generated trajectories.
    \end{itemize}

The remainder of this paper is organized as follows. Section \ref{sec3} formally defines the problem. Section \ref{sec32} introduces the two trajectory planning algorithms. Section \ref{sec4} presents the vehicle dynamics model and details the NMPC strategy. Numerical results demonstrating the effectiveness of the proposed approach are shown in Section \ref{sec5}, followed by concluding remarks in Section \ref{sec6}.

\section{Problem Formulation}\label{sec3}
We consider the problem of jointly planning trajectories for $N$ vehicles at an intersection. The objective is to minimize time for all vehicles to pass the intersection while avoiding collisions and respecting constraints on vehicle motions. We consider general intersections whose structure dictates and provides predetermined paths to follow. The knowledge of the paths allows to identify potential intersection/collision points in advance. By knowing these collision points, confined areas around them are considered conflict areas where only one vehicle may pass at a time. We will address this problem by considering a restricted problem in a high-dimensional configuration space, see Section~\ref{2b}. However, in order to do so, we first define some preliminary definitions. 

\subsection{Preliminaries}\label{sec3s1}
Each Vehicle $i$ in the intersection has a path $\pmb{\phi}_i(s)$ where $s \in [0, \bar{s}_i]$ and $\pmb{\phi}(s) \in \mathbb{R}^2 = \mathcal{S}$ for each $s \in [0, \bar s_i]$; $\mathcal{S}$ represents the $\textit{physical space}$. All paths are represented by $\{\pmb{\phi}_i(s_i)\}_{i=1}^N$. The paths defined by the $\pmb{\phi}_i$'s are assumed arc-length parameterized.  Since those paths represent road paths in the intersection, they are assumed to be possible to follow and not possess unfavorable characteristics such as self-intersecting, and intersecting with each other many times, except when two vehicles travel on the same path with different start points and/or end/final points. We define the \textit{path configuration space} (or simply configuration space) $\mathcal{C} = [0, \bar{s}_1] \times [0, \bar{s}_2] \times ... \times [0, \bar{s}_N]$ in which $\times$ is the Cartesian product. Note here that $[0, \bar{s}_1]$ is an interval of the real line and hence a set. If we write, for example, $[1,2]^\top$, such an object is a column vector containing the two elements $1$ and $2$. The notation should be apparent by the context. For the $N$ vehicles, each following their respective path, we denote the positions of the vehicles (which are functions of time) along their paths by $\pmb{s} = [s_1, s_2, \ldots, s_N]^\top \in \mathcal{C}$ and the velocities along the vehicle-paths by $\pmb{v} = \dot{\pmb{s}} = [v_1, v_2, \ldots, v_N]^\top \in \mathbb{R}^N$. The 2-dimensional velocity in the physical space for each vehicle $i$ is $\pmb\phi_i(v_i)$.

An intersection point (also referred to as collision or conflict point) is a tuple $(\pmb{s}_{\text{ip}}, i,j, s_{\text{ip},i},s_{\text{ip},j})$ for which the following holds. For Vehicle $i$ and Vehicle $j$, $s_{\text{ip},i} \in [0, \bar{s}_{i}]$, and $s_{\text{ip},j} \in [0, \bar{s}_{j}]$, $\pmb{p}_{\text{ip}} = \pmb{\phi}_{i}(s_{\text{ip},i}) = \pmb{\phi}_{j}(s_{\text{ip},j})$. The point $\pmb{p}_{\text{ip}} \in \mathcal{S}$ is referred to as the physical location of the intersection point. 
 
Note that with this definition of intersection point, there may be more than one intersection point with the same physical location $\pmb{p}_{\text{ip}}$. However, each intersection point corresponds to a unique pair of vehicles. Besides $\pmb{p}_{\text{ip}}$, we define for convenience $\pmb{s}_{\text{ip}} = [s_{\text{ip},i}, s_{\text{ip},j}]^\top$. 

A conflict set or region in the physical space $\mathcal{S}$ for an intersection point is a set of points near the physical location of an intersection point in the form of a compact convex set (e.g., a disk or a rectangle) containing the physical location of the intersection point in its interior and such that each of the two paths in physical space of the two vehicles enter and exit only once.

We enumerate the intersection points for each pair of vehicles $(i,j)$. Assume $(\pmb{s}_{\text{ip}}, i,j, s_{\text{ip},i},s_{\text{ip},j})$ is the $k$'th intersection point for the vehicle pair $(i,j)$. We define the conflict region $\mathcal{O}_{ij}^{(k)} \subset \mathcal{S}$ as follows. We begin by constructing a rectangular region as a subset of 
$[0, \bar s_i] \times [0, \bar s_j]$, that corresponds to the conflict region in $\mathcal{S}$. 
This rectangular region is defined as  $\pmb{s}_{\text{ip}} + [\delta_{ij,1}^{(k)}, \delta_{ij,2}^{(k)}] \times [\delta_{ij,3}^{(k)}, \delta_{ij,4}^{(k)}]$, where $\delta_{ij,1}^{(k)}$ is the length along the path $\pmb{\phi}_i$ between the point where the path $\pmb{\phi}_i$ enters the conflict region in $\mathcal{S}$ and the point $\pmb{s}_{\text{ip}}$; $\delta_{ij,2}^{(k)}$ is the length along the path $\pmb{\phi}_i$ between $\pmb{s}_{\text{ip}}$ and the point where the the path $\pmb{\phi}_i$ leaves the conflict region in $\mathcal{S}$; and $\delta_{ij,3}^{(k)}$ and $\delta_{ij,4}^{(k)}$ are equivalently defined for vehicle $j$. 
From the rectangular region, we now define $\mathcal{O}^{(k)} \subset \mathcal{S}$ as the set of vectors in $\mathcal{C}$ for which the $i$'th element is in the set $s_{\text{ip},i} + [\delta_{ij,1}^{(k)}, \delta_{ij,2}^{(k)}]$, the $j$'th element is in the set $s_{\text{ip},j} + [\delta_{ij,3}^{(k)}, \delta_{ij,4}^{(k)}]$ and all the remaining $N-2$ elements are equal to $0$.

The full set of conflict regions $\mathcal{O}$ is obtained by taking the union over all such pairwise conflict regions, where
\begin{equation}
\mathcal{O} = \bigcup_{i < j} \bigcup_{k} \mathcal{O}_{ij}^{(k)}
\end{equation}

\begin{figure}[t]
    \centering
    \subfloat[Intersection scenario for 3 vehicles. Conflict regions are red circles.\label{fig:3v}]{
        \includegraphics[width=0.7\linewidth]{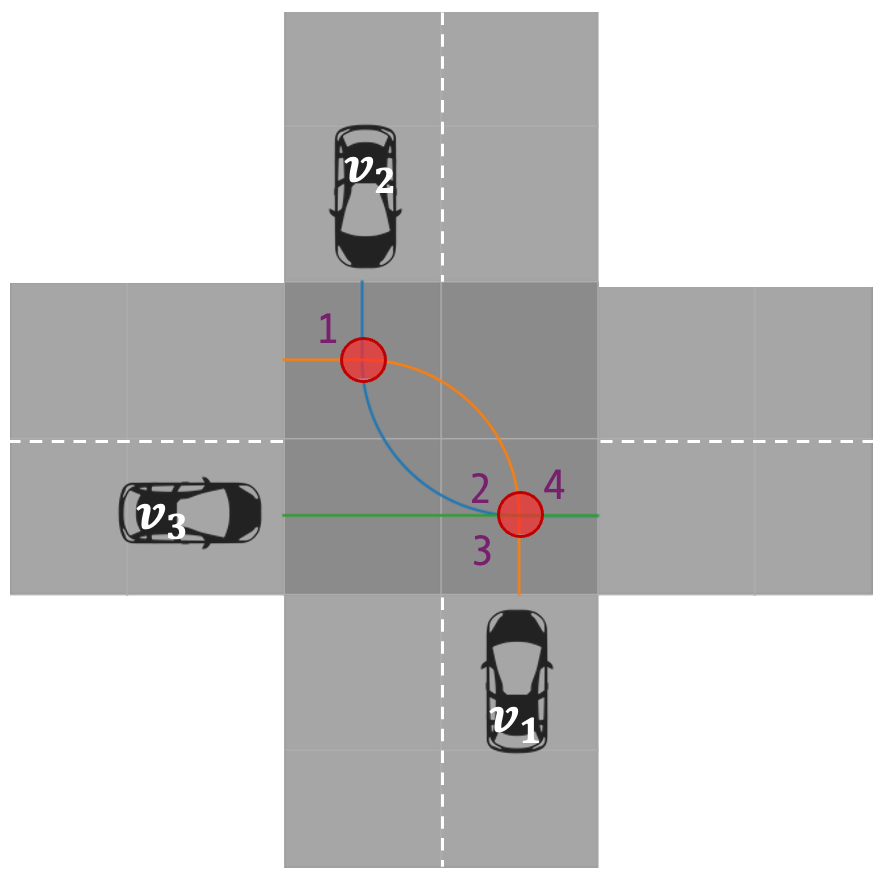}
    }

    \subfloat[Intersection scenario for 2 vehicles on the same path. Conflict regions are red circles.\label{fig:2v}]{
        \includegraphics[width=0.7\linewidth]{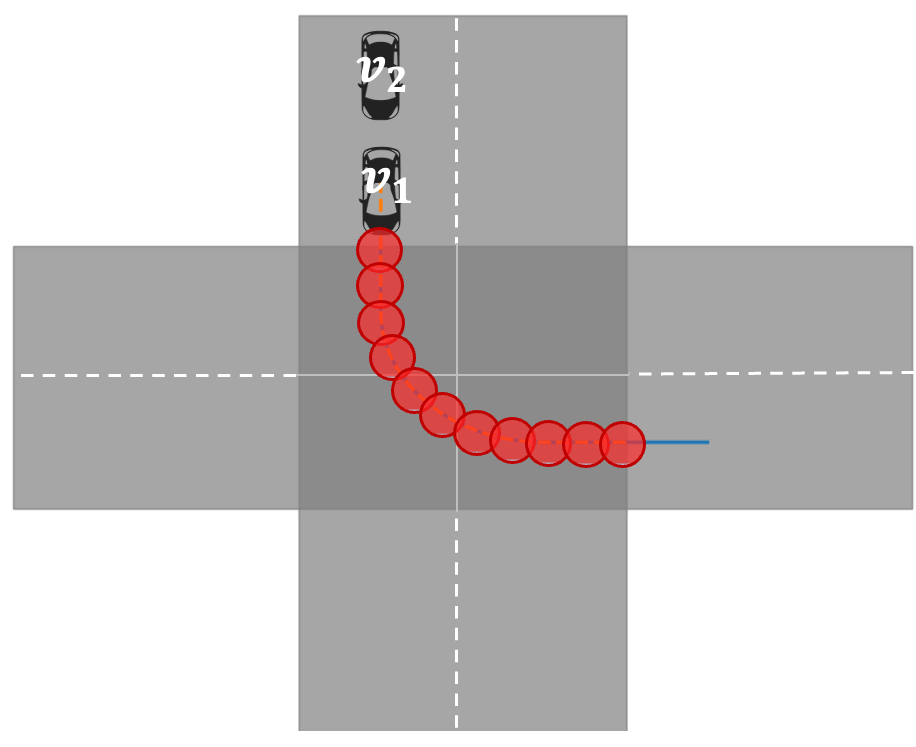}
    }
    \caption{Intersection scenarios with conflict regions shown in red.}
    \label{fig:scens}
\end{figure}

\textit{Example 1:} Consider the intersection scenario illustrated in Figure \ref{fig:3v}, where we have four conflict points with four conflict regions in the physical space $\mathcal{S}$ that are discs. These conflict regions should in practice be chosen larger to ensure no collisions, but are here chosen small for illustration purposes. 

\begin{figure}[t]
    \centering
    \includegraphics[width=\linewidth]{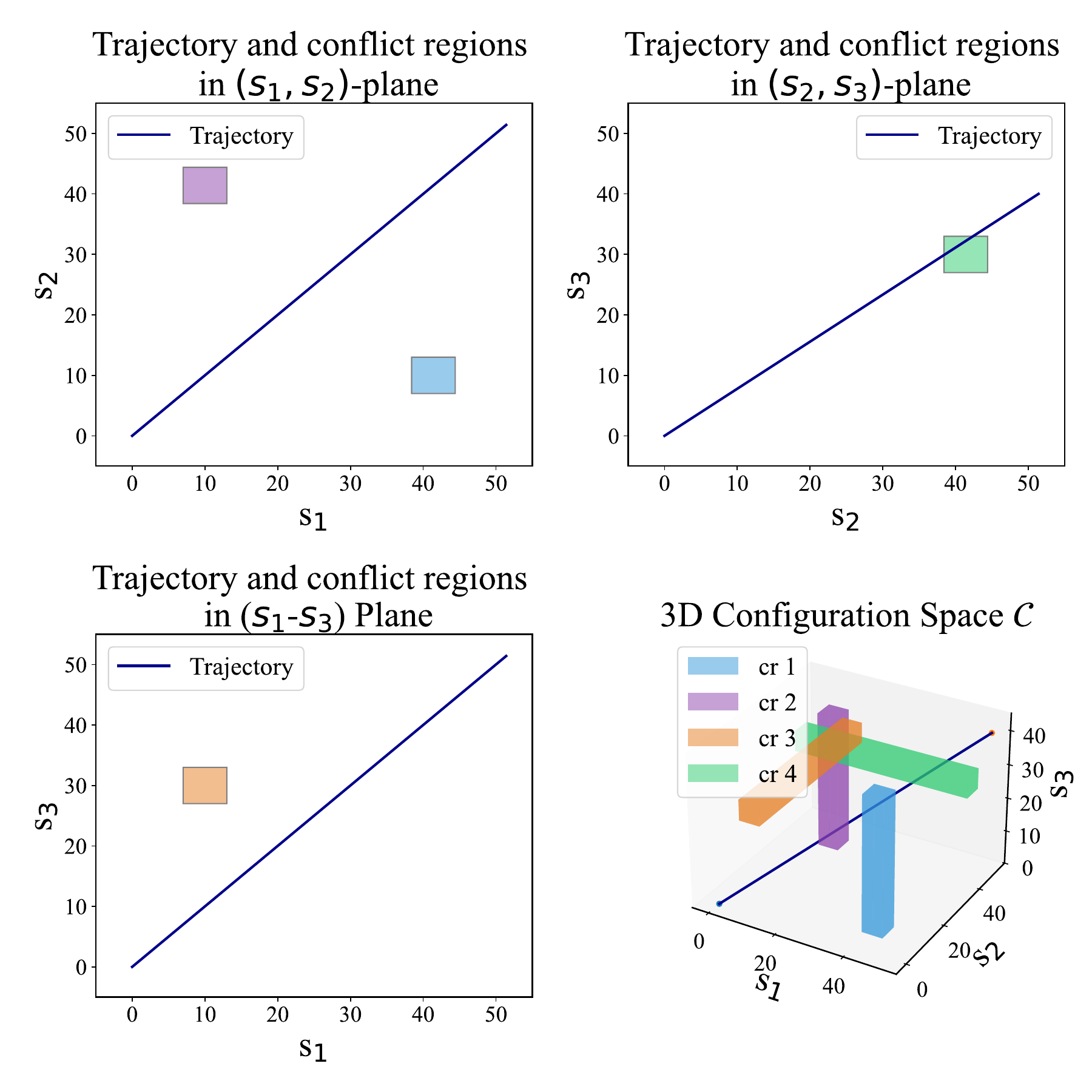}
    \caption{Configuration Space ($\mathcal{C}$) and subspaces thereof for vehicles with constant velocity.}
    \label{fig:position}
\end{figure}

The path configuration space or simply configuration space $\mathcal{C}$ for this scenario is illustrated in the bottom-right sub-figure of Figure~\ref{fig:position}.  In this sub-figure, the green point, $[0,0,0]^\top$ and the red point, $[\bar{s}_1, \bar{s}_2, \bar{s}_3]^\top$ is the initial point for the vehicles in $\mathcal{C}$ and the final point for the vehicles in $\mathcal{C}$, respectively. 
The blue, the purple, the orange, and green N-dimensional polyhedral are the conflict regions in $\mathcal{C}$, i.e.,  the $\mathcal{O}_{ij}^{(k)}$'s. 
In this example, if all vehicles travel at a certain constant speed, vehicle 2 and vehicle 3 pass through a conflict region at the same time. The positions of the vehicles are illustrated in Figure \ref{fig:position}. The main challenge lies in avoiding this phenomenon, i.e., to make sure that no more than one vehicle is in a conflict region at any given time.

\begin{figure}[t]
    \centering
    \includegraphics[width=\linewidth]{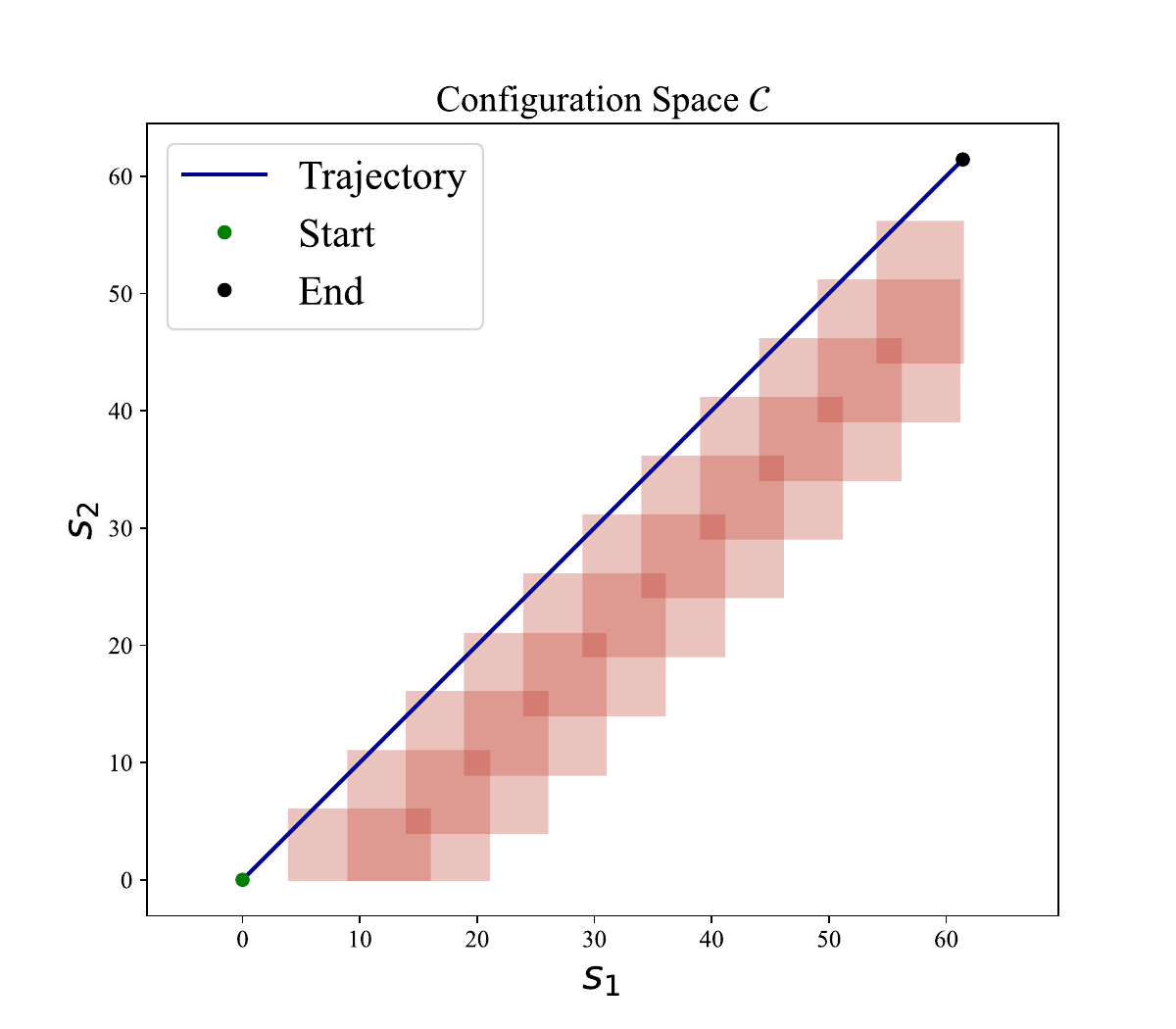}
    \caption{Configuration space ($\mathcal{C}$) of vehicles for the scenario in Figure \ref{fig:2v}. In example 2, we have 2 vehicles; therefore, the configuration space is two-dimensional. }
    \label{fig:2obstacles}
\end{figure}

\textit{Example 2:} In the example, illustrated in Figure \ref{fig:2v}, two vehicles move along the same path, and we model the configuration space by placing a sequence of evenly spaced conflict regions along the path. Again, these conflict regions should in practice be chosen larger to ensure no collisions, but are here chosen small for illustration purposes. For each conflict region disc in $\mathcal{S}$ there is a collision if the two vehicles are positioned inside at the same time. As shown in Figure~\ref{fig:2obstacles}, if both vehicles move at a certain constant speed, their trajectories pass through all conflict regions, resulting in collisions.

\subsection{Problem Formulation}\label{2b}

Now we formulate the configuration-space trajectory planning problem. The problem is to
find the minimum-length, in the Euclidean distance sense, collision-free trajectory $\pmb{\gamma}: [0, 1] \rightarrow \mathcal{C}$ from the start position $[0,0, \ldots, 0]^\top$ to the goal position $\pmb{\bar{s}} = [\bar{s}_1, \bar{s}_2, \ldots, \bar{s}_N]^\top$ in $\mathcal{C}$. Under the restriction that $\|\pmb{v}(t)\|_2$ is constant (or static) over time, where $\pmb{v}(t) = \dot{\pmb{s}(t)} = [v_1(t), v_2(t), \ldots, v_N(t)]^\top$, such a trajectory in $\mathcal{C}$ corresponds to choosing velocities for the vehicles
such that all the vehicles reach their final positions along their respective path together in the physical space in minimum (transportation) time. One can think of this as moving all vehicles safely from an initial safe constellation through the intersection to a final safe desired constellation, in minimum time. 
The described trajectory planning problem in $\mathcal{C}$ may be equivalently formulated as follows:
\begin{equation}
\begin{aligned}
\min_{\pmb{\gamma}} \quad & \int_0^1 \|\dot{\pmb{\gamma}}(\tau)\|_2 \, d\tau \\
\text{s.t.} \quad & \pmb{\gamma}(\tau) \in \mathcal{C} \setminus \mathcal{O}, \quad \forall \tau \in [0, 1], \\
& \dot{\pmb{\gamma}}(\tau) \geq \pmb{0}, \quad \forall \tau \in [0, 1], \\
& \pmb{\gamma}(0) = \pmb{0}, \quad \pmb{\gamma}(1) = \pmb{\bar{s}},
\end{aligned}
\label{eq:optimal_path}
\end{equation}
where $\dot{\pmb{\gamma}}(\tau) \geq \pmb{0}$ enforces element-wise non-negativity of $\dot{\pmb{\gamma}}(\tau)$, which ensures monotonic progression along each vehicle's path.
Here we assume $\tau$ is normalized to be in the interval $[0, 1]$, which is not necessary, and any other positive length interval for $\tau$ provides an equivalent solution. The ratio between a pair of elements in $\dot{\pmb{\gamma}}(\tau)$ provides the ratio between the velocities for the corresponding pair of vehicles at the normalized time $\tau$.

\section{High-dimensional Trajectory Planning by 2D-decomposition. }\label{sec32}

Here we address Problem~\eqref{eq:optimal_path} by using decomposition approaches, where we solve (N-1) number of 2D-graph search problems optimally to obtain a near-optimal feasible solution to Problem~\eqref{eq:optimal_path}. Two such decomposition approaches are presented. The first, the incremental approach, uses the solution from the $i$'th 2D-problem to solve the $(i+1)$'th 2D-problem, where $i \in \{1,2, \ldots, N-2\}$. For the second, the pairwise planning (or simply pairwise) approach, independent pairwise solutions are combined iteratively. 

\subsection{Incremental Approach}

This approach is iterative and solves $(N-1)$ 2D-problems in a sequence. 
For the first 2D-problem, two vehicles are selected, Vehicle $i_1$ and Vehicle $i_2$, and a restricted problem is considered where the number of vehicles is 2. The path configuration space $\mathcal{C}^{(1)}$ for this problem is two-dimensional and conflict regions in $\mathcal{C}^{(1)}$ are rectangles. 
A DAG visibility graph is constructed where reachable corner points of rectangle conflict regions in $\mathcal{C}^{(1)}$ as well as the start position and final position comprise nodes. The shortest path problem in the DAG visibility graph is solved via graph search to find the optimal conflict-free trajectory $\pmb{\gamma}_1$ in $\mathcal{C}^{(1)}$, of length $\|{\pmb{\gamma}}_1\|_2 = \int_0^1 \|{\dot{\pmb{\gamma}}_1}(\tau)\|_2 \, d\tau$. $\pmb{\gamma}_1$ comprises a few connected line-segments, which means that the computation of $\|\pmb{\gamma}_1\|_2$ is fast and explicit and does not require numeric integration. 

Any subsequent 2D-problem is constructed and solved as follows. We create a new 2D space, where one of the two dimensions corresponds to a previously not considered vehicle, and the other dimension corresponds to the previously considered vehicles. 
For example, for the second 2D-problem to be solved, we consider a new vehicle $i_3 \not\in\{i_1, i_2\}$. Now there are three vehicles considered in total, but the configuration space $\mathcal{C}^{(2)}$ we construct is two-dimensional. From before we know $s_{i_3}$ is the variable that defines the position of vehicle $i_3$ along the corresponding axis in $\mathcal{C}^{(2)}$ and $\pmb{\phi}_{i_3}(s_{i_3})$, defines the corresponding position in physical space. 

We represent the positions of vehicle $i_1$ and vehicle $i_2$ by one variable, call it $s_{i_1,i_2}(\tau) = \int_0^\tau \|{\dot{\pmb{\gamma}}_1}(\tilde \tau)\|_2 \, d\tilde \tau$ that is in the interval $[0, \|\pmb{\gamma}_1\|_2]$. It captures the position along $\pmb{\gamma}_1$. As earlier stated, $\pmb{\gamma}_1$ comprises a few connected line segments and $s_{i_1,i_2}(\tau)$ can be explicitly computed without the need for numeric integration. 
Now, for a given $s_{i_1,i_2}$, we get the corresponding positions of vehicles $i_1$ and $i_2$ as follows.  $\pmb{\phi}_{i_1}([\pmb{\gamma}_1(s_{i_1,i_2})]_1)$ and $\pmb{\phi}_{i_2}([\pmb{\gamma}_1(s_{i_1,i_2})]_2)$ is the position of Vehicle $i_1$ in $\mathcal{S}$ and the position of Vehicle $i_2$ in $\mathcal{S}$, respectively, where $[\cdot]_l$ is notation for the $l$'th element of a vector. Now, the shortest path problem in the two-dimensional $\mathcal{C}^{(2)}$ is solved by finding the optimal conflict-free trajectory in $\mathcal{C}^{(2)}$ via graph search. Conflict regions in this in the space $\mathcal{C}^{(2)}$ are rectangles obtained via projection onto $\pmb{\gamma}_1$. 

More generally, for any of the 2D problems in the sequence of 2D problems, conflict regions are rectangles and optimal solutions are obtained by solving shortest path problems via visibility DAG-graphs. The final conflict-free trajectory comprises a trajectory in $\mathcal{C}$.  The overall approach is formally presented in Algorithm \ref{alg1}.

\begin{algorithm}
\caption{Incremental Approach}
\label{alg1}
\begin{algorithmic}[1]
\State \textbf{Input:} vehicle paths $\{\pmb{\phi}_i\}_{i=1}^n$, conflict region $\mathcal{O} \subset [0, \bar{s}_1] \times \cdots \times [0, \bar{s}_n]$
\State \textbf{Output:} conflict-free trajectory $\pmb{\gamma}: [0,1] \to [0, \bar{s}_1] \times \cdots \times [0, \bar{s}_n]$
\State $\mathcal{V} \gets \{1, \dots, n\}$
\State Select two vehicles $i, j \in \mathcal{V}$
\State $\mathcal{C'} \gets [0, \bar{s}_{i}] \times [0, \bar{s}_{j}]$
\State $\pmb{\gamma} \gets \mathrm{search2D}(i, j, \mathcal{O}_{{c'}})$
\State $\mathcal{V} \gets \mathcal{V} \setminus \{i, j\}$
\While{$\mathcal{V} \neq \emptyset$}
    \State Select $r \in \mathcal{V}$
    \State $\mathcal{C'} \gets [0, \bar{s}_r] \times [0, \|\pmb{\gamma}\|_2]$
    \State $\mathcal{O}_{\mathcal{C'}} \gets \mathrm{proj}_{\mathcal{C'}}(\mathcal{O})$ \; 
    \State $\pmb{\gamma}' \gets \mathrm{search2D}(r, \boldsymbol{\gamma}, \mathcal{O}_{C'})$
    \State $\pmb{\gamma} \gets \mathrm{lift}(\pmb{\gamma}, \pmb{\gamma}')$
    \State $\mathcal{V} \gets \mathcal{V} \setminus \{r\}$
\EndWhile
\State \Return $\pmb{\gamma}$
\end{algorithmic}
\end{algorithm}

In Algorithm~\ref{alg1}, the $\mathrm{proj}_{\mathcal{C}'}$-function is used to extract the two-dimensional rectangle conflict regions in $\mathcal{C}'$ from the high-dimensional configuration space. More formally, given the full conflict set $\mathcal{O} \subset \mathcal{S}$ and a the two-dimensional $\mathcal{C}'$, the projection $\mathrm{proj}_{\mathcal{C}'}(\mathcal{O})$ returns the conflict regions $\mathcal{O}_{\mathcal{C}'}$ in  $\mathcal{C}'$.

The function $\mathrm{search2D}(i, j, \mathcal{O}_{\mathcal{C}'})$ returns a collision-free trajectory in $\mathcal{C}'$ (that avoids all rectangles in $\mathcal{O}_{\mathcal{C}'}$) by solving the shortest path problem in a weighted directed graph (DAG) $\mathcal{G} = (\mathcal{N}, \mathcal{E})$, where $\mathcal{N}$ is the set of nodes and $\mathcal{E} \subseteq \mathcal{N} \times \mathcal{N}$ is the set of edges. Each node $n \in \mathcal{N}$ corresponds to a 2D position $\bold{p}_n = [x_n, y_n]^T \in \mathbb{R}^2$, and each edge $e = (u, v) \in \mathcal{E}$ has a weight $w_e = \|\bold{p}_u - \bold{p}_v\|_2$ equal to the Euclidean distance between its 2D endpoints $\bold{p}_u$ and $\bold{p}_v$ corresponding to $u$ and $v$, respectively. 

The function $\mathrm{lift}(\pmb{\gamma}, \pmb{\gamma}')$ 
extends the trajectory $\pmb{\gamma}$ to a trajectory in a space with one more dimension by using the newly constructed 2D trajectory $\pmb{\gamma}'$. Each sample of $\pmb{\gamma}'$ is on the form $[s_r,\ell]^\top$, where $\ell$ denotes an arc-length coordinate along $\pmb{\gamma}$. The value $\ell$ is used to obtain the corresponding point or vector $\pmb{\gamma}(\ell)$, which is efficiently computed since $\pmb{\gamma}$ comprises connected line-segments. The lifted higher-dimensional vector or point corresponding to $[s_r, \ell]^\top$ is $[s_r, \pmb{\gamma}(\ell)]^\top$. 

We see that the order of selection of vehicles in Algorithm~\ref{alg1} is not specified, i.e., how $i$ and $j$ are selected at line 4 and how each new $r$ is selected at line 9. For each of the selections of vehicles, the algorithm provides a final $\pmb{\gamma}$-trajectory in $\mathcal{C}$. Each order can equivalently be represented by a permutation (of the $N$ vehicles). The number of permutations is $N!$. However, the 
resulting trajectory is independent of the order of $(i,j)$ at line 4 (the same trajectory is obtained for $(i,j)$ and $(j,i)$). Thus, the number of permutations to be evaluated is $\frac{N!}{2}$, which becomes intractable for large $N$. However, as will be seen in Section~\ref{sec5}, using Algorithm~\ref{alg1} with only one randomly selected permutation (or order) provides very good results in general. Nevertheless, we could also introduce additional structure that restricts the number of permutations needed, at the expense of restricting the number of possible $\pmb{\gamma}$-trajectories that can be generated. Such a procedure is introduced in the next subsection.

\subsection{Pairwise Approach}
In this approach the number of permutations needed for generating all possible trajectories is smaller than that for the incremental approach.  The key idea is to organize the $N$ vehicles or equivalently the $N$ vehicle paths into a binary tree. 

In the first pass of the method, we select $\lfloor \frac{N}{2} \rfloor$ disjoint pairs of vehicles.
If $N$ is odd, one vehicle remains unpaired. We then run 2D graph to compute a joint $\pmb{\gamma}$-trajectory for each $(i,j)$-pair. 
For the second pass, for each $(p,q)$-pair of joint trajectories $(\pmb{\gamma}_p, \pmb{\gamma}_q)$, we construct a new coordination space $\mathcal{C} = [0, \|\pmb{\gamma}_p\|_2] \times [0, \|\pmb{\gamma}_q\|_2]$ and perform 2D graph to compute a joint $\pmb{\gamma}$-trajectory. If the number of resulting $\pmb{\gamma}$’s is odd, the unpaired vehicle or trajectory is carried forward to the next pass. 

This process continues until a single $\pmb{\gamma}$ remains, which represents the final trajectory.
For instance, when $N=4$ we form two pairs and produce two $\pmb{\gamma}$'s in the first pass, then for those two trajectories, we compute the final $\pmb{\gamma}$-trajectory in the second pass (which represents the trajectory for all four vehicles). If $N=5$, in the first pass, pair four vehicles are included as two pairs and one is left aside; in the second pass, the two generated $\pmb{\gamma}$'s are combined into a trajectory and the same vehicle that was left aside in the first pass is still left aside; and in the final, third, pass there is only one pair used for computation of the final $\pmb{\gamma}$.

Although this binary tree decomposition approach does not change the number of 2D-problems needed to compute a final $\pmb{\gamma}$ in comparison to Algorithm~\ref{alg1} (which is still $(N-1)$ many), it greatly reduces the number of distinct permutations to check, see Proposition \ref{pro1} below.

\begin{proposition} \label{pro1}
For Algorithm~\ref{alg2}, any $\pmb{\gamma}$-trajectory obtained for a permutation or ordering of vehicles can equivalently be obtained by a permutation in a set containing 

\begin{equation}
    \frac{N!}{\left| \operatorname{Aut}(T_N) \right|}
\end{equation}
where $\left|\operatorname{Aut}(T_{N})\right| = 2^{\sum_{j=1}^{m}(2^{k_{j}}-1)}$.
Here $\sum_{j=1}^{m} 2^{k_{j}} = N$ (the number of vehicles) and the 
values of $m$ and $k_{j}$ are determined directly by $N$ and indicate
how many symmetric subtrees appear at each level of the coordination tree.\cite{harary2014graphical}.
\end{proposition}


\begin{proof}
In the pairwise planning algorithm, coordination proceeds by recursively pairing agents and merging their joint trajectories. This structure can be modeled as a binary tree with $N$ leaves, each representing one vehicle. Many permutations of the leaves (vehicle orderings) produce equivalent binary trees, known as automorphic trees.

Each symmetric subtree of depth $ k_j $ contains $ 2^{k_j} $ leaves and $ 2^{k_j} - 1 $ internal nodes. Each internal node can swap its left and right branches, contributing a factor of 2 to the number of automorphisms. Hence, the total number of automorphisms for one such subtree is $ 2^{2^{k_j} - 1} $, and for all subtrees combined:
\begin{equation}\label{eq:aut}
    \left| \operatorname{Aut}(T_N) \right| = \prod_{j=1}^m 2^{2^{k_j} - 1} = 2^{\sum_{j=1}^m (2^{k_j} - 1)}.
\end{equation}
By dividing the total number of permutations $N!$ by the expression in \eqref{eq:aut}, we obtain the number of distinct coordination trees we need to evaluate.
\end{proof}

\begin{algorithm}
\caption{Pairwise Approach}
\label{alg2}
\begin{algorithmic}[1]
\State \textbf{Input:} Vehicle paths $\{\pmb{\phi}_i\}_{i=1}^n$, conflict region $\mathcal{O} \subset [0, \bar{s}_1] \times \cdots \times [0, \bar{s}_n]$
\State \textbf{Output:} Conflict-free trajectory $\pmb{\gamma}: [0,1] \to [0, \bar{s}_1] \times \cdots \times [0, \bar{s}_n]$
\State $\mathcal{V} \gets \{1, \dots, n\}$
\While{$|\mathcal{V}| > 1$}
    \State Initialize $\mathcal{V}_{\text{next}} \gets \emptyset$
    \ForAll{disjoint pairs $(i, j)$ in $\mathcal{V}$}
        \State $\mathcal{C} \gets [0, \bar{s}_i] \times [0, \bar{s}_j]$
        \State $\mathcal{O}_{c} \gets \mathrm{proj}_{\mathcal{C}}(\mathcal{O})$ \; 
        \State $\pmb{\gamma}_{ij} \gets \mathrm{search2D}(i, j, \mathcal{O}_{c})$
        \State Append $\pmb{\gamma}_{ij}$ to $\mathcal{V}_{\text{next}}$
    \EndFor
    \If{$|\mathcal{V}|$ is odd}
        \State Carry the unpaired element to $\mathcal{V}_{\text{next}}$
    \EndIf
    \State $\mathcal{V} \gets \mathcal{V}_{\text{next}}$
\EndWhile
\State $\pmb{\gamma} \gets \mathcal{V}$ \; \% Final trajectory in reduced space
\State $\pmb{\gamma} \gets \mathrm{lift}(\pmb{\gamma}, \{\pmb{\phi}_i\}_{i=1}^n)$ \; \% Final trajectory in $\mathcal{C}$
\State \Return $\pmb{\gamma}$
\end{algorithmic}
\end{algorithm}

In Algorithm \ref{alg2} the function $\mathrm{lift}(\pmb{\gamma}, \{\pmb{\phi}_i\}_{i=1}^n)$ maps the two-dimensional trajectory into the full N-dimensional configuration space $\mathcal{C} = [0, \bar{s}_1] \times \cdots \times [0, \bar{s}_n]$. The operations $\mathrm{proj}$ and $\mathrm{search2D}(i, j, \mathcal{O}_{\mathcal{C}})$ retain the same definitions as described in Algorithm~\ref{alg1}.

\begin{figure}
  \centering

  \begin{subfigure}[b]{\linewidth}
    \centering
    \includegraphics[width=.88\linewidth]{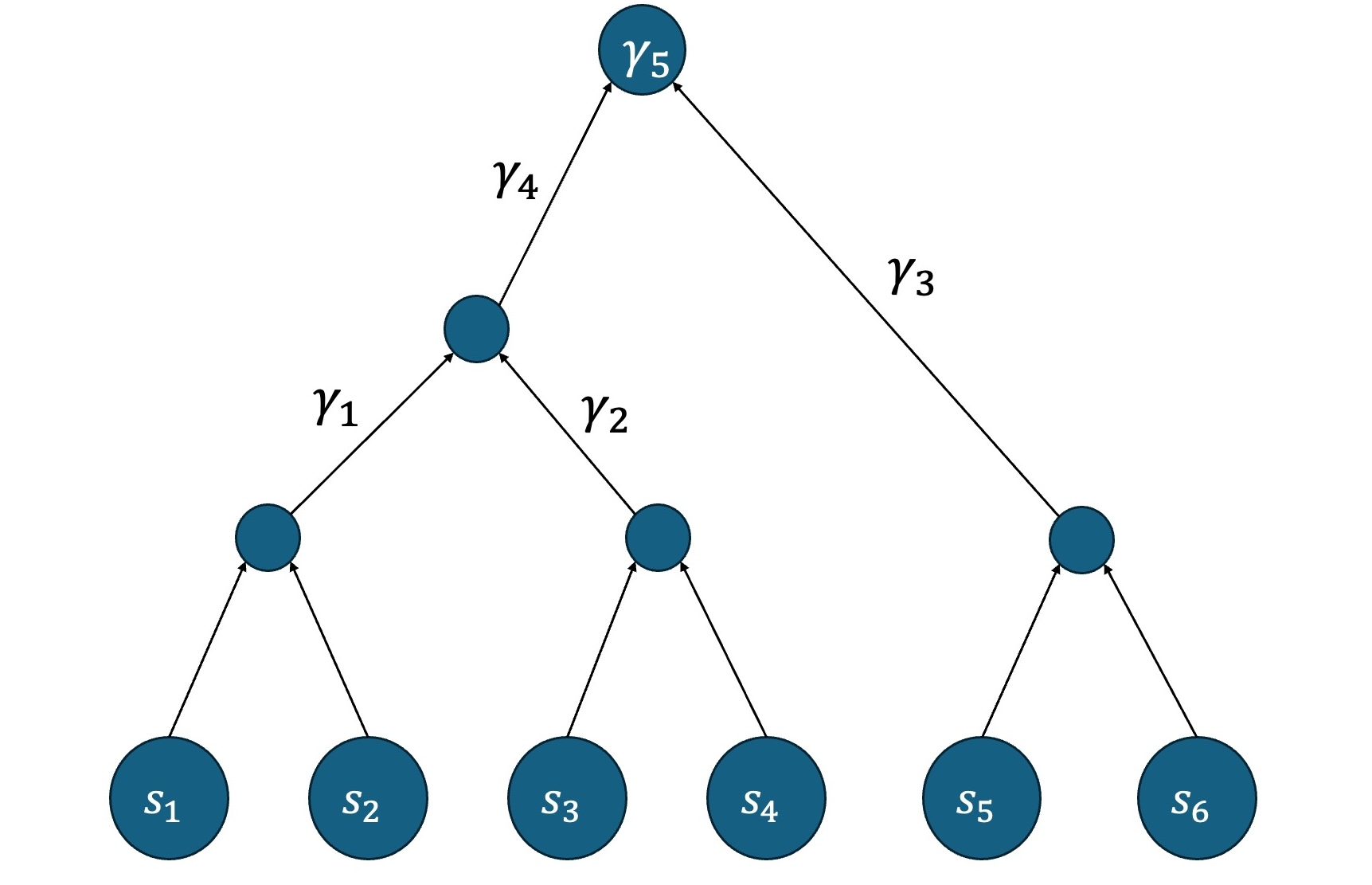}
    \caption{6 vehicles}
    \label{fig:gamma6}
  \end{subfigure}
  \hfill
  \begin{subfigure}[b]{\linewidth}
    \centering
    \includegraphics[width=0.9\linewidth]{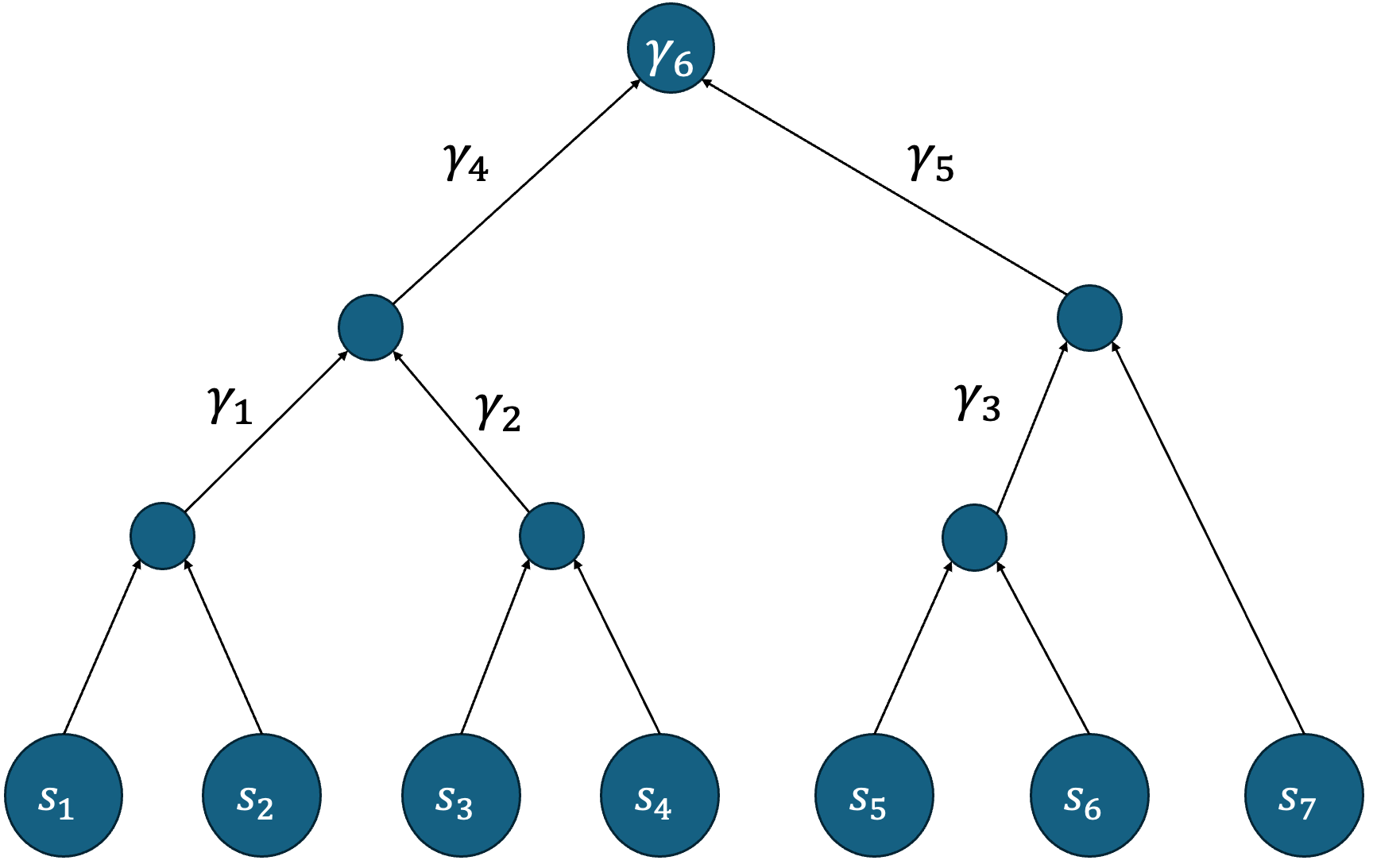}
    \caption{7 vehicles}
    \label{fig:gamma7}
  \end{subfigure}
  \hfill
  \begin{subfigure}[b]{\linewidth}
    \centering
    \includegraphics[width=\linewidth]{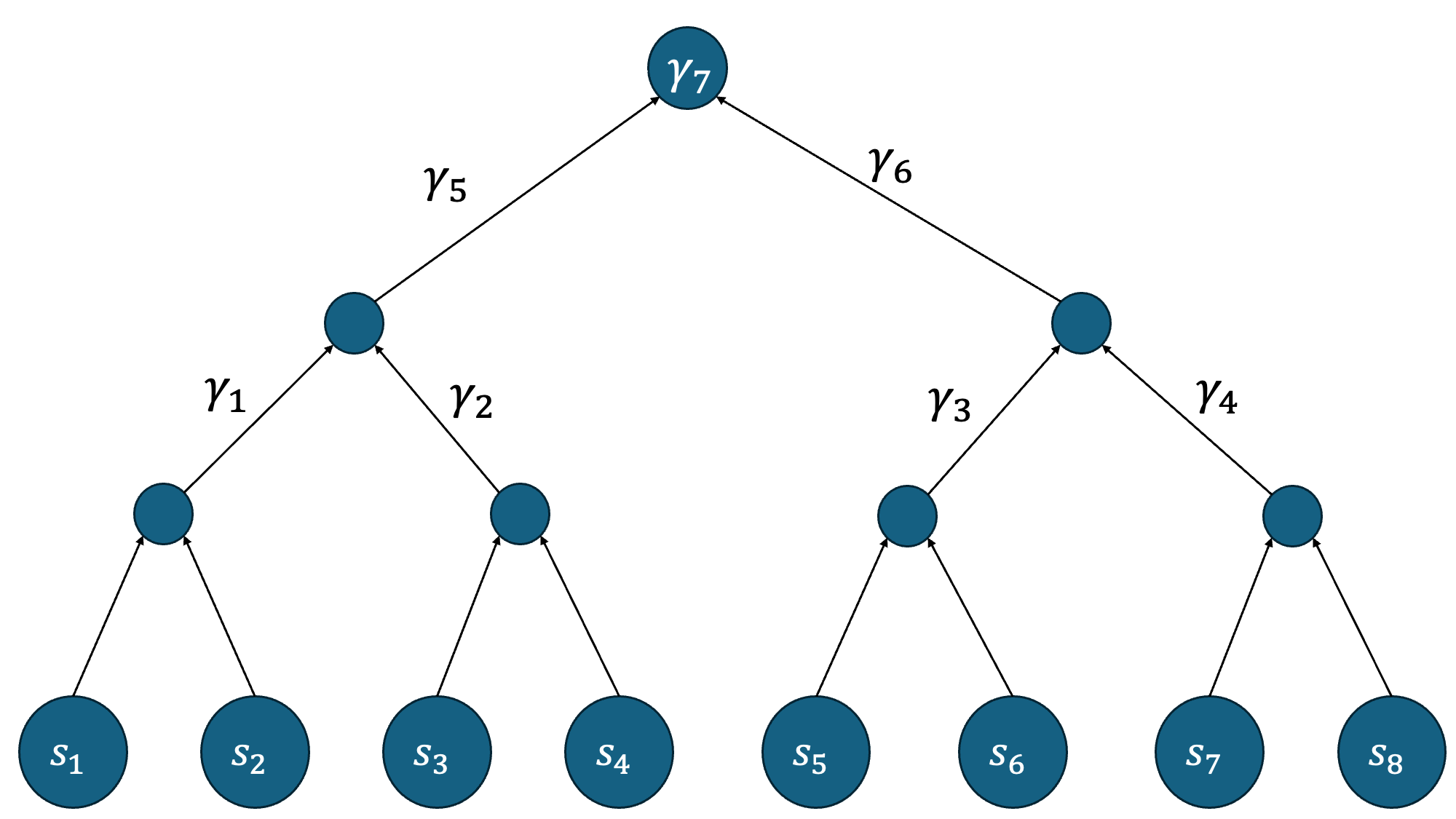}
    \caption{8 vehicles}
    \label{fig:gamma8}
  \end{subfigure}
  \caption{Illustration of the algorithm's behavior for 6, 7, and 8 vehicles. The figure shows how the pairwise approach constructs the high-dimensional configuration space $\mathcal{C}$.}
  \label{fig:gamma-comparison}
\end{figure}

Figure \ref{fig:gamma-comparison} demonstrates the application of Algorithm \ref{alg2} for three different scenarios involving 6, 7, and 8 vehicles, respectively. For each scenario, the structure is shown as a binary tree, where each leaf node represents a vehicle.

We now analyze the computational complexity of the proposed planning algorithms based on the number of agents and the number of conflict points.

\subsection{Analysis of the Algorithms Connection to Vehicle-Control}
In this section we provide some on computational complexity and feasibility of the solutions generated by Algorithm~\ref{alg1} and Algorithm~\ref{alg2}. 

\begin{proposition}\label{prop:comp:complex}
The computational complexity for both Algorithm~\ref{alg1} and Algorithm~\ref{alg2} is
\begin{equation}
    \mathcal{O}\big(N r^2),
\end{equation}
where $r$ be the total number of conflict points.
\end{proposition}

\begin{proof}
For both algorithms, $N-1$ 2D problems are solved. For each such 2D problem, at most one call of each of the functions $\mathrm{proj}_{\mathcal{C}}$, $\mathrm{lift}$, and $\mathrm{search2D}$ is involved. 

For $\mathrm{search2D}$, each conflict region contributes to at most four nodes (corners of rectangle obstacles). The visibility DAG comprises at most $4r + 2$ nodes (where the start-position and end position is included), and fewer than $(4r + 2)^2$ directed edges. Since obtaining the shortest path in the DAG is $\mathcal{O}(|\mathcal{E}|) = \mathcal{O}({r^2})$, where $|\mathcal{E}|$ is the number of edges, $\mathrm{search2D}$ is $\mathcal{O}({r^2})$. Any $\pmb{\gamma}$-trajectory constructed at any stage comprises at most $\mathcal{O}(r)$ connected line-segments and both $\mathrm{proj}_{\mathcal{C}}$ and $\mathrm{lift}$ is $\mathcal{O}(r)$. So the total computational complexity is $\mathcal{O}(Nr^2)$

\end{proof}
It should be said that in practice the visibility graphs for the 2D-problems are sparse and far from fully connected. So the provided result in Proposition~\ref{prop:comp:complex} should be seen as conservative. 

Now, we turn to the important question of feasibility of the solution. Suppose $\pmb{\gamma}$ is obtained by either Algorithm~\ref{alg1} or Algorithm~\ref{alg2}. By feasible solution, we mean that $\pmb{\gamma}$ is such that for each time, and any conflict region, there is no pair of vehicles for which both of the two vehicles in the pair are inside the conflict region. Formally, if we normalize time to the interval $[0,1]$ so that $\tilde{\pmb{\gamma}}(\tau) = \pmb{\gamma}(\frac{\tau}{\|\pmb{\gamma}\|_2})$,  then the constraints in \eqref{eq:optimal_path} must be fulfilled for $\tilde{\pmb{\gamma}}(\tau)$.

\begin{proposition}\label{pro_g}
If $\pmb{\gamma}$ is obtained by either Algorithm~\ref{alg1} or Algorithm~\ref{alg2}, $\pmb{\gamma}$ is a feasible solution (no collisions occur).  

\end{proposition}

\begin{proof}
We use an induction argument to show the desired result. 
First, when only two vehicles $i$ and $j$ are considered, the 2D search is done directly in $[0,\bar{s}_i] \times [0,\bar{s}_j]$. All obstacles $\mathcal{O}_{ij}^{(k)}$ in the two-dimensional space appear as rectangles, and the returned trajectory explicitly avoids them. So there are no collisions for such $\pmb{\gamma}$-trajectories.
Secondly, suppose solve a 2D problem for a pair of collision-free trajectories or a vehicle and a collision-free trajectory. This covers the situation in the iterations for either Algorithm~\ref{alg1} or Algorithm~\ref{alg2}, and for simplicity we refer for both situations to a pair of trajectories. The projected conflict regions in this new two-dimensional space comprise rectangles. The solution to the 2D problem is a safe two-dimensional trajectory (that avoids all rectangles). This two-dimensional trajectory comprise connected line-segments. After lifting, each of the two previous trajectories have been (positively) rescaled (for each line-segment), but are still safe under such scaling. By induction, the final $\pmb{\gamma}$-trajectory is safe. Monotonicity of the solution (i.e., that vehicles move forward along their paths) is ensured by the structure of the constructed visibility graphs. 

\end{proof} 

As previously stated,  $\pmb{\gamma}$ comprises connected line-segments. This means that $\frac{\dot{\pmb{\gamma}}}{\|\pmb{\gamma}\|_2} = \frac{\pmb{v}}{\|\pmb{v}\|_2}$ is a piece-wise constant (in time) unit-vector. This piece-wise constant unit vector provides the relative velocities between vehicles (whereas the absolute velocities can be changed to incorporate velocity constraints stemming from the specific intersection scenario or the dynamics of vehicles). This piece-wise constant unit vector will be incorporated in the cost function for the vehicle-controllers in the next section and, as such, provides the connection between trajectory planning and the control.

\section{Vehicle Dynamics and Model Predictive Control}\label{sec4}
\label{sec:controller}

In this section, we evaluate the planner introduced in the previous section within a nonlinear MPC framework. Specifically, we use a nonlinear model to assess trajectory tracking and control, framing the task as a finite-horizon optimal control problem solved with an NMPC approach. The vehicle dynamics are represented as \cite{rajamani2011vehicle}

\begin{equation} \label{eq:model}
    \mathbf{x}_{k+1} = f(\mathbf{x}_{k}, \mathbf{u}_{k})
\end{equation}
Here, the vehicle is modeled with the state and control input vectors defined as
$$
\mathbf{x}_k = \begin{bmatrix}
x_k \\ y_k \\ \psi_k \\ v_k \\ \delta_k
\end{bmatrix}, \qquad
\mathbf{u}_k = \begin{bmatrix} a_k \\ \dot\delta_k \end{bmatrix},
$$
and discrete-time update equations
\begin{align}
x_{k+1} &= x_k + v_k \cos\psi_k\,\Delta t, \\
y_{k+1} &= y_k + v_k \sin\psi_k\,\Delta t, \\
\psi_{k+1} &= \psi_k + \frac{v_k}{L} \tan\delta_k\,\Delta t, \\
v_{k+1} &= v_k + a_k\,\Delta t, \\
\delta_{k+1} &= \delta_k + \dot\delta_k\,\Delta t,
\end{align}

In these equations $x_k, y_k$ denote the rear-axle position, $\psi_k$ is the yaw (heading) angle, $v_k$ is the longitudinal speed, $\delta_k$ is the front-wheel steering angle, $a_k$ is the longitudinal acceleration, $\dot\delta_k$ is the steering rate,  $L$ is the wheelbase, and $\Delta t$ is the discretization time step.

At each time $t$, we solve the following finite-horizon optimal problem
\begin{equation}
\begin{split}
\min_{\{\mathbf{u}_{k|t}\}_{k=t}^{N-1}} \;
\sum_{k=t}^{t+N-1} &
    \left[
        (\mathbf{x}_{k|t} - \mathbf{x}_{k|t}^{\rm ref})^\top Q (\mathbf{x}_{k|t} - \mathbf{x}_{k|t}^{\rm ref})\right.\\
      &+ \mathbf{u}_{k|t}^\top R \mathbf{u}_{k|t}\\
      &\left.+ (\mathbf{u}_{k|t} - \mathbf{u}_{k-1|t})^\top S (\mathbf{u}_{k|t} - \mathbf{u}_{k-1|t})
    \right] \\
&  
+ (\mathbf{x}_{N|t} - \mathbf{x}_{N|t}^{\rm ref})^\top Q_f (\mathbf{x}_{N|t} - \mathbf{x}_{N|t}^{\rm ref}) \\[1.2em]
\textrm{s.t.} \quad
& \mathbf{x}_{k+1|t} = f(\mathbf{x}_{k|t}, \mathbf{u}_{k|t}), \: k=t,\dots,t+N-1, \\[0.5em]
& \mathbf{u}_{k|t} \in \mathcal{U}, \quad \mathbf{x}_{k|t} \in \mathcal{X}, \: k=t,\dots,t+N-1, \\[0.5em]
& \mathbf{x}_{0|t} = \mathbf{x}(t).
\end{split}
\end{equation}

In this equation, $\mathbf{x}_{k|t}$ represents the predicted state of each agent at future time step $k$, computed from the current state $\mathbf{x}(t)$ using the dynamic model in Equation~\ref{eq:model}. The reference state $\mathbf{x}^{\mathrm{ref}}{k|t}$ is obtained from a precomputed geometric path, parameterized by arc length. 
Progress along the arc-length is guided by a reference velocity defined as $ v_{k|t}^{\rm ref} = \alpha_{k|t} v_{\max}$ where $ v_{\max}$ denotes the maximum allowable cruising velocity. The scaling factor $\alpha_{k|t} \in (0,1]$ is obtained from the piecewise constant unit velocity vector generated by the trajectory planner. This unit vector is rescaled by dividing by its maximum component across all vehicles at each planning step, ensuring that at least one vehicle attains the full speed $v_{\max}$, while the remaining vehicles follow with proportionally scaled velocities. In this way, the relative motion prescribed by the trajectory planner is preserved, and all vehicles progress consistently while avoiding conflict regions.

The cost function is defined using weight matrices $Q \geq 0$, $R \geq 0$, $S \geq 0$, and $Q_f \geq 0$, which are block-diagonal but selected to be diagonal in practice to penalize deviations from the reference trajectory. The vehicle dynamics $f(\mathbf{x}_k, \mathbf{u}_k)$ are subject to constraints that will be detailed in the following 
$$
\mathcal{U} = \left\{ \mathbf{u} :
    a_{\min} \leq a_k \leq a_{\max}, \quad
    \dot{\delta}_{\min} \leq \dot{\delta}_k \leq \dot{\delta}_{\max}
\right\},
$$
$$
\mathcal{X} = \left\{ \mathbf{x} :
    \delta_{\min} \leq \delta_k \leq \delta_{\max}, \quad
    a_{\rm lat}^{\min} \leq \frac{v_k^2}{L}\tan\delta_k \leq a_{\rm lat}^{\max}
\right\}.
$$

The MPC problem is solved at each time step using the Sequential Least Squares Programming (SLSQP) algorithm. A warm-start strategy is applied by initializing the solver with the previous control sequence.

\section{Numerical Results}\label{sec5}
In this section, we evaluate the proposed algorithms for trajectory planning with or without subsequent NMPC under various conditions.
First, we evaluate the two trajectory-planning algorithms in three scenarios involving 6, 7, and 8 vehicles 
Secondly, we benchmark our trajectory planning to problem-adapted form of the lower-bound MILP method proposed in \cite{7987071} used for a real-world intersection scenario with 20 vehicles. Finally, we focus on this real-world scenario with 20 vehicles and analyze the performance of the NMPC.

\subsection{Evaluation of Proposed Trajectory-Planning Algorithms}

\begin{figure*}
  \centering
\begin{subfigure}[b]{0.32\textwidth}
    \includegraphics[width=\linewidth]{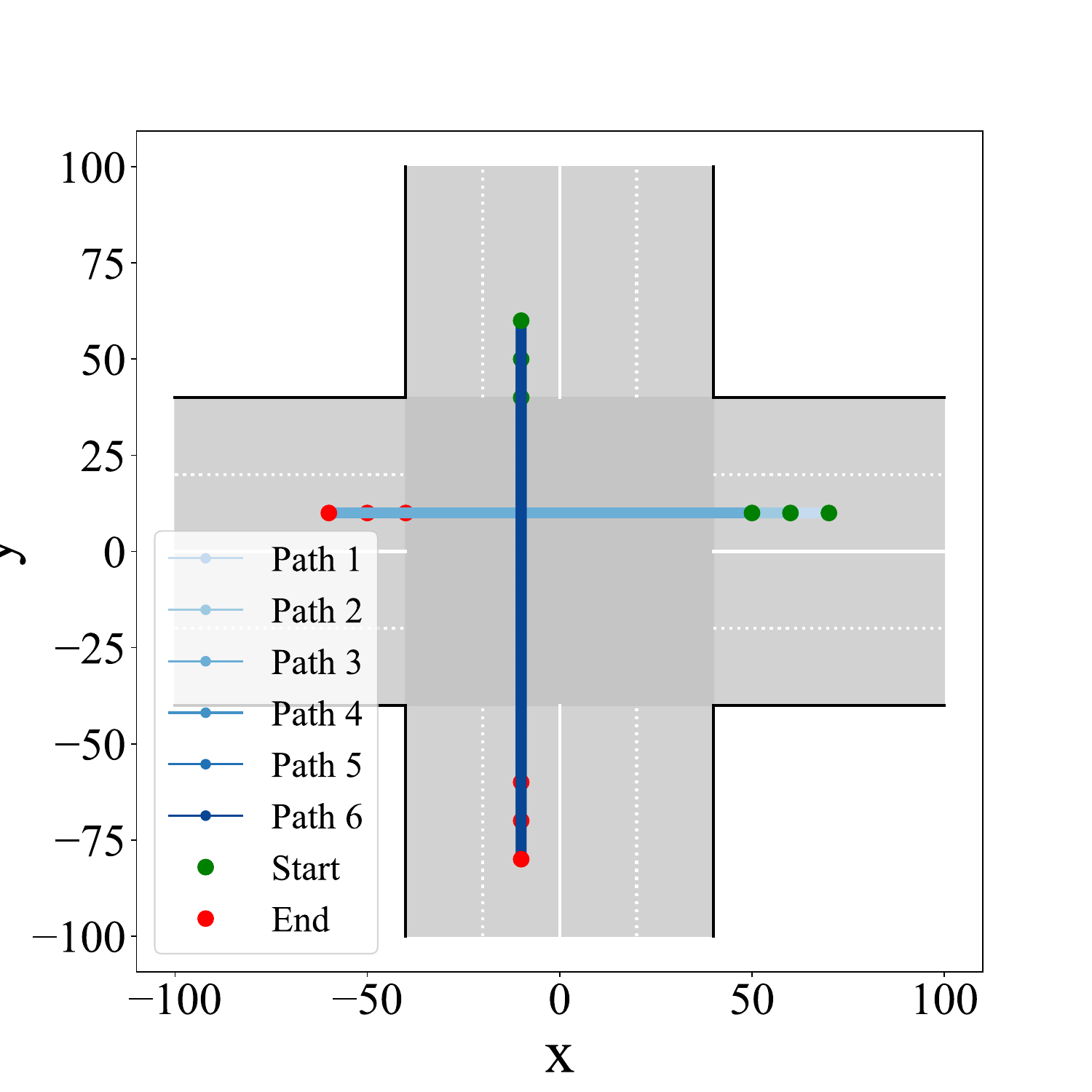}
    \caption{6 vehicles.}
    \label{fig:v6}
\end{subfigure}
  \hfill
  \begin{subfigure}[b]{0.32\textwidth}
    \includegraphics[width=\linewidth]{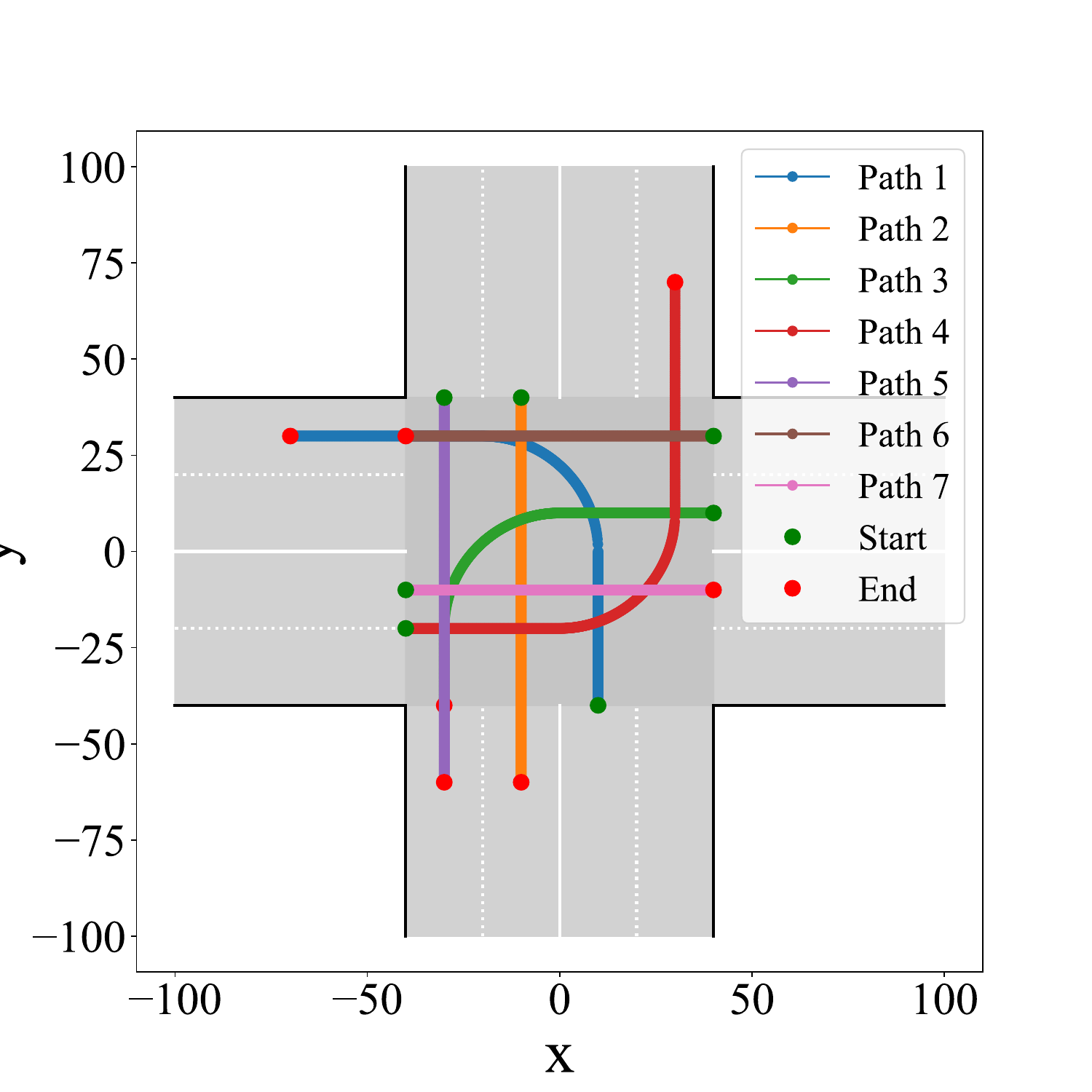}
    \caption{7 vehicles.}
    \label{fig:v7}
  \end{subfigure}
  \hfill
  \begin{subfigure}[b]{0.32\textwidth}
    \includegraphics[width=\linewidth]{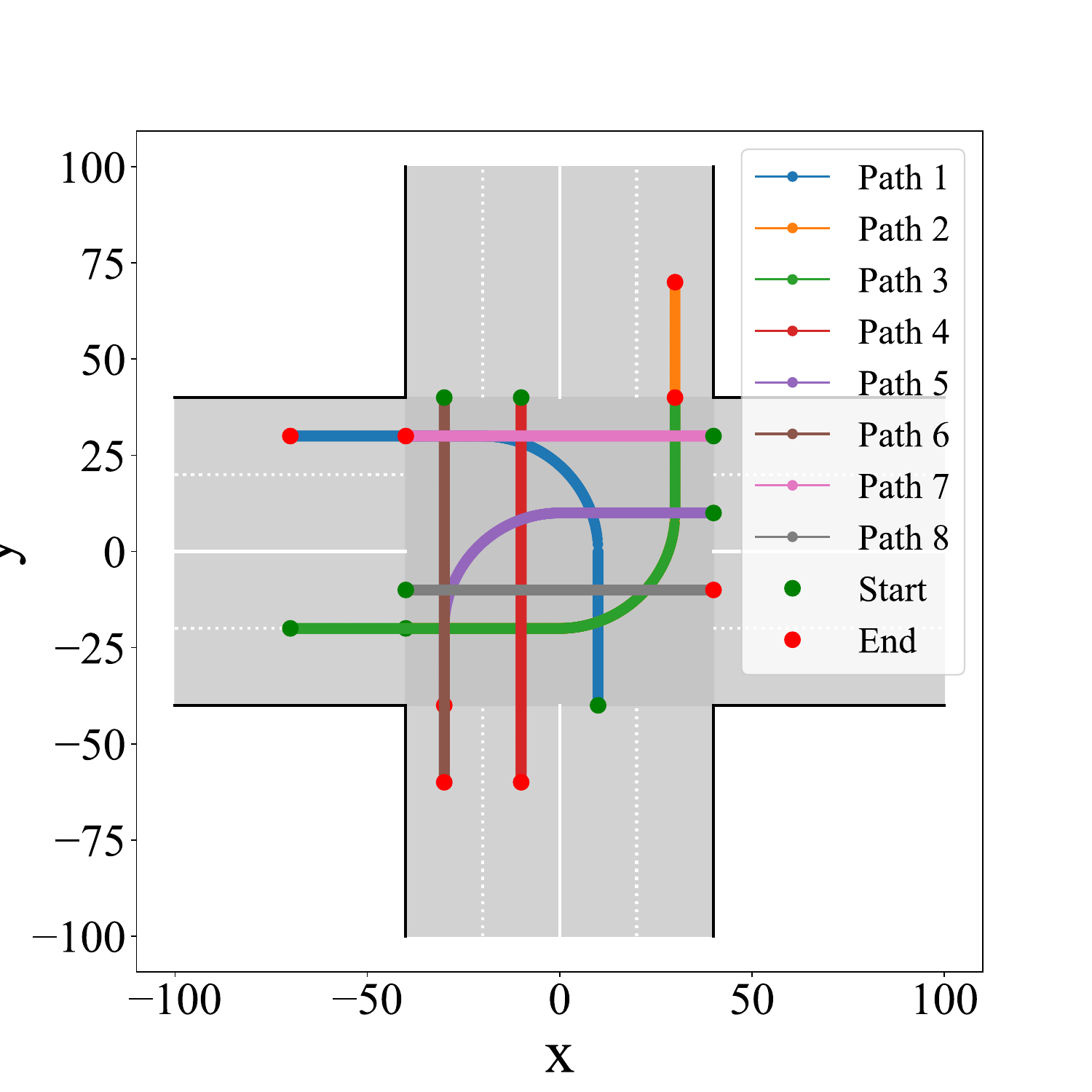}
    \caption{8 vehicles.}
    \label{fig:v8}
  \end{subfigure}
  \caption{Three intersection scenarios with varying vehicle counts. (a) Six vehicles are traveling on two paths (three vehicles per path). (b) Seven paths with one vehicle on each. (c) Eight vehicles on seven paths, where one path carries two vehicles and each remaining path carries one.}
  \label{fig:different scenarios}
\end{figure*}

Here we focus on comparing the two planning algorithms, Algorithm \ref{alg1} and Algorithm \ref{alg2}. The focus is on trajectories in the configuration space $\mathcal{C}$ generated by either one of the two algorithms. These trajectories serve as suboptimal solutions to the optimization problem \ref{eq:optimal_path} presented in Section \ref{2b}. The straight-line path from start to goal position in the N-dimensional configuration space $C$ provides a useful lower bound for performance evaluation. However, it is in general not a feasible trajectory, i.e., it intersects $\mathcal{O}$. But, we use it as a baseline to assess the quality of the suboptimal trajectories produced by our algorithms, i.e., to get an upper bound on the gap to optimality. 

We compare Algorithm \ref{alg1} and Algorithm \ref{alg2} across three distinct scenarios, depicted in Figure \ref{fig:different scenarios}. In the first scenario (Figure \ref{fig:v6}), six vehicles travel along two shared paths, highlighting that the algorithms can handle scenarios with multiple vehicles on the same path. In the second scenario (Figure \ref{fig:v7}), there are seven vehicles on seven independent paths (i.e., there is one vehicle per path). In the third scenario (Figure \ref{fig:v8}), there is a mixture of the situation in the first scenario and the situation on the second scenario: seven paths are used by eight vehicles. 

\begin{figure}
  \centering
\begin{subfigure}[b]{\linewidth}
    \includegraphics[width=\linewidth]{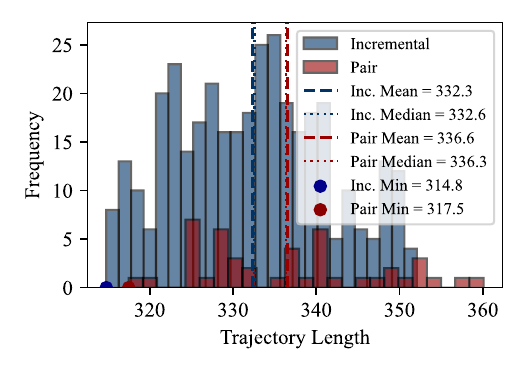}
    \caption{6 vehicles.}
    \label{fig:rv6}
\end{subfigure}
  \hfill
  \begin{subfigure}[b]{\linewidth} \centering
    \includegraphics[width=\linewidth]{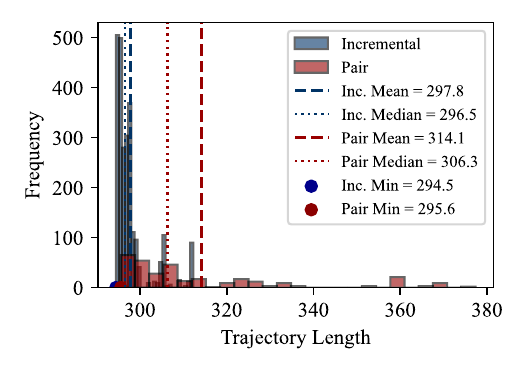}
    \caption{7 Vehicles.}
    \label{fig:rv7}
  \end{subfigure}
  \hfill
  \begin{subfigure}[b]{\linewidth } \centering
    \includegraphics[width=\linewidth]{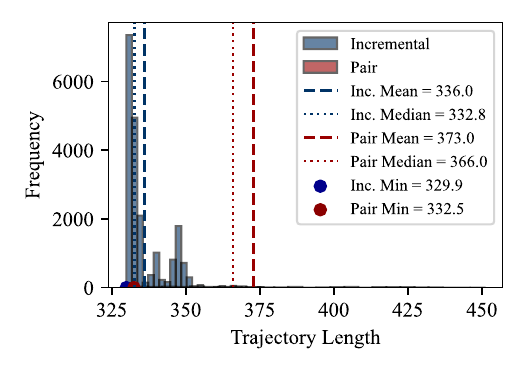}
    \caption{8 vehicles.}
    \label{fig:rv8}
  \end{subfigure}
  \caption{Comparison of length of trajectory for the Incremental (blue) and Pair (red) algorithms across all permutations in three traffic scenarios: (a) six vehicles on two shared paths, (b) seven vehicles each on its own path, and (c) eight vehicles on seven paths (one path carrying two vehicles). Solid vertical lines mark each algorithm’s mean length, dashed lines its median, and colored dots the minimum length achieved.}
  \label{fig:res 3 scenarios}
\end{figure}

A shorter path in $\mathcal{C}$ corresponds to a shorter overall transportation time, i.e., the time to move the vehicles from their initial positions in the physical space to the goal (or final) positions. 
Results for all three scenarios for all possible permutations are summarized in Table~\ref{tab:shortest_line_comparison} and Figure \ref{fig:res 3 scenarios}. Column ``Lower bound'' displays the lower bound on the optimal solution tothe  optimization problem \eqref{eq:optimal_path}, obtained by computing the Euclidean distance between the start point and the end (goal) point in $\mathcal{C}$.  

In Scenario 1, since there are multiple conflict points along shared paths,  
which are on the straight-line trajectory in 6D space due to vehicles sharing just two routes, 
it is to be expected that both algorithms yield longer trajectories relative straight line comprising the lower bound. However, in Scenarios 2 and 3, where the paths are distinct or more distinct, the trajectory lengths computed by both algorithms are much closer to the lower bound, i.e., the upper bound on the optimality gap is small. 
As shown in Figure \ref{fig:res 3 scenarios}, Algorithm \ref{alg1} consistently produces shorter trajectories-- consequently faster transportation times--than Algorithm \ref{alg2} in every scenario. Conversely, Table~\ref{tab:runtimes} shows that Algorithm \ref{alg2} is $74.6\%$, $75.73\%$, and $97.5\%$ faster than Algorithm \ref{alg1} for scenarios 1, 2 and 3, respectively.

\begin{table}
  \centering
  \begin{tabular}{@{} l ccc @{}}
    \toprule
    \multirow{2}{*}{\textbf{Scenario}} 
      & \multicolumn{3}{c}{\textbf{Shortest path length}} \\ 
    \cmidrule(l){2-4}
      & \textbf{Incremental Alg.} & \textbf{Pair Alg.} & \textbf{Lower Bound} \\ 
    \midrule
    Scenario 1 (6 vehicles) & 314.79 & 317.48 & 283.01 \\ 
    Scenario 2 (7 vehicles) & 294.45 & 295.58 & 291.07 \\ 
    Scenario 3 (8 vehicles) & 329.94 & 332.54 & 326.14 \\ 
    \bottomrule
  \end{tabular}
  \caption{Comparison of the shortest paths obtained by the proposed algorithms with lower bound Euclidean distance (i.e., straight line distance) between start and goal positions across the three scenarios.}
  \label{tab:shortest_line_comparison}
\end{table}

\begin{table}
  \begin{tabularx}{\linewidth}{@{}l *{2}{>{\centering\arraybackslash}X}@{}}
    \toprule
    \multirow{2}{*}{\textbf{Scenario}} 
      & \multicolumn{2}{c}{\textbf{Runtime (seconds)}} \\ 
    \cmidrule(l){2-3}
      & \textbf{Incremental Alg.} & \textbf{Pair Alg.} \\ 
    \midrule
    Scenario 1 (6 vehicles) & 45.19 & 11.47 \\ 
    Scenario 2 (7 vehicles) & 9.52 & 2.31 \\ 
    Scenario 3 (8 vehicles) & 603.62 & 14.87 \\ 
    \bottomrule
  \end{tabularx}
  \caption{Runtimes for Alg. \ref{alg1} vs.\ Alg. \ref{alg2} for the three scenarios}
  \label{tab:runtimes}
\end{table}

\begin{figure}
    \centering
    \includegraphics[width=\linewidth]{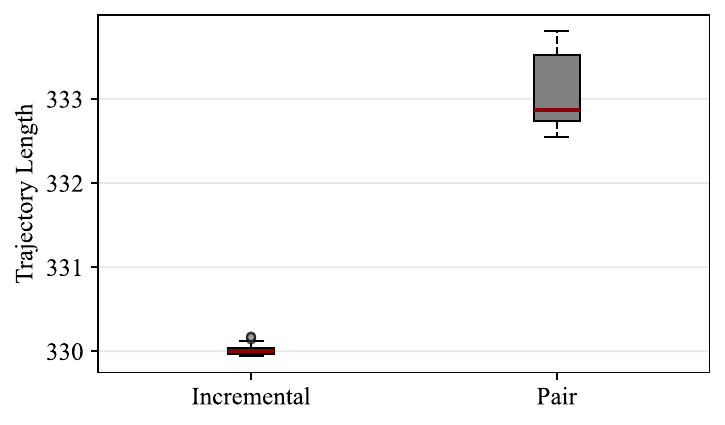}
    \caption{Comparison of the minimum trajectory length found by each algorithm after 5 seconds. Distribution over 30 independent repetitions.}
    \label{fig:comparing_random}
\end{figure}

Since the previous results alone do not clearly indicate which algorithm performs better overall, we adopted an alternative comparison method. In this approach, we constrained the computation time to 5 seconds and ran 
each algorithm for the Scenario 3 with 8 vehicles (Figure \ref{fig:v8}),
using random permutations of the vehicle ordering. We did it 30 times to reduce the impact of randomness and ensure fairness. In each run, we recorded the shortest trajectory (i.e., the best result) produced within that time limit. In Figure~\ref{fig:comparing_random} the distribution the length of the 30 best trajectories is displayed. First of all, Algorithm~\ref{alg1} was faster; it was able to compute about 170 trajectories in 5 seconds, while Algorithm \ref{alg2} produced roughly 93 trajectories. 
The distribution for the incremental algorithm (Algorithm \ref{alg1}) is compact and close to 330. In contrast, the pairwise algorithm (Algorithm \ref{alg2}) is centered closer to 333 and stretches, and is wider. In plain terms, Algorithm \ref{alg1} not only consistently produces shorter trajectories in the 5-second periods, but it does so with very little variation from run to run. 
This suggests that, given the same time budget, Algorithm \ref{alg1} is both faster and more reliable at producing a short trajectory. 
When the number of vehicles is small, for example smaller than six, the number of permutations to evaluate for using the pairwise approach is small, and the algorithm can be evaluated for all quickly. Furthermore, if each vehicle follows a distinct path, experiments seem to suggest that the performance of the pairwise algorithm is good. In scenarios outside this setting, especially for problems involving large $N$, the incremental algorithm, combined with a randomized permutation of the vehicle ordering, is expected to perform well.

\subsection{Benchmark Setup Using Time-schedule Method}
\begin{figure}
    \centering
\includegraphics[width=0.9\linewidth]{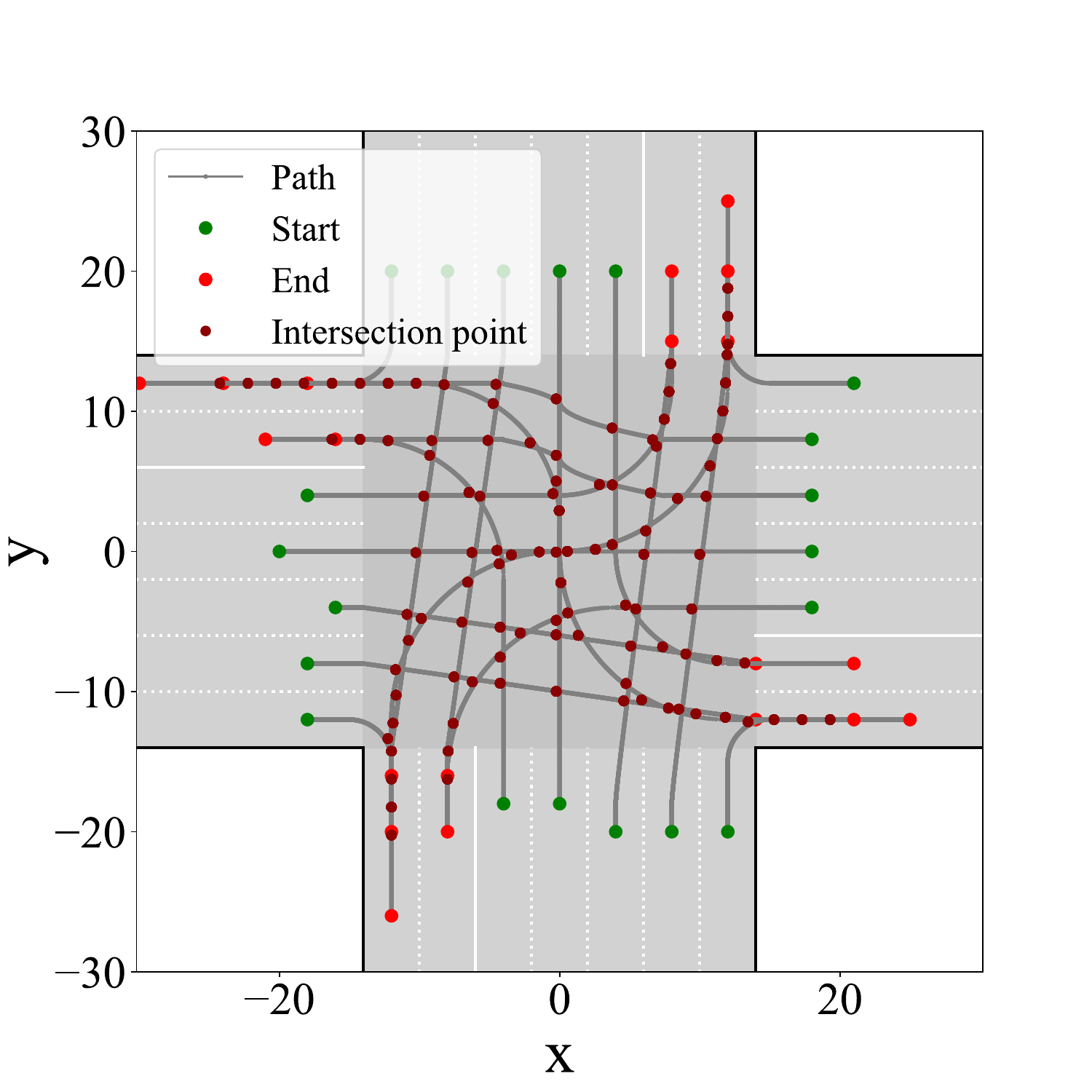}
    \caption{Real-world scenario involving 20 vehicles.}
    \label{fig:rw_scen}
\end{figure}

In this section, we evaluate our approach for a scenario that more closely reflects real-world conditions, and compare the performance to a time-schedule optimization method. For this, we employ the intersection layout presented in \cite{7987071}, illustrated in Figure \ref{fig:rw_scen}. For benchmarking, we use an adapted mixed integer linear programming (MILP) formulation via the lower-bound formulation in the same paper. This adaptation provides the best trajectory that this scheduling-based approach can generate in the setting considered in this paper, allowing a direct comparison with the result of our proposed method. For more details about the job-shop scheduling framework, readers can refer to \cite{7987071}. The method from this paper is chosen for comparison because its framework aligns closely with our problem setting and is widely used in subsequent works.

We compared the makespan, i.e., transportation time, and runtime of the MILP-based approach with our method. To reduce randomness, we used the same maximum vehicle velocity in both algorithms and ran each method 1000 times with random permutations for the incremental approach. The MILP was solved using the Gurobi optimizer, a highly reliable and well-established solver \cite{gurobi}. 
In contrast, our incremental approach relies on our own implementation, which has not optimized, see Section~\ref{sec:imp:det}.
Table \ref{tab:comparison} reports the average makespan, i.e., transportation time, and runtime.

The experimental results demonstrate that the proposed incremental approach provides better performance in both computational efficiency and solution quality. First, our method is approximately 41 times faster (in runtime). This is due to the inherent difference in algorithmic complexity. The MILP formulation translates the conflict resolution into an NP-Hard combinatorial problem that requires the solver to navigate a discrete search space defined by binary variables to determine the optimal agent ordering. In contrast, our approach uses efficient 2D graph-searching for the shortest path, which also provides the order for the vehicles shall pass conflict regions. Secondly, the incremental algorithm achieves a coordination makespan, or transportation time, that is over $52\%$ shorter than that of the MILP-based approach. This reduction in makespan is attributed to the difference in method for managing overlapping conflict regions. The MILP's job-shop framework enforces a strict mutual exclusion constraint, treating each segment of conflict as a distinct resource that must be processed sequentially. This leads to an artificially long and conservative trajectory, as vehicles are forced to wait even when sufficient clearance exists. Our method, operating in the continuous configuration space, aggregates these conflicts into a unified geometric obstacle. This enables the solver to find a continuous trajectory that reduces transportation time while guaranteeing safety.

\begin{table}[h!]
\centering
\begin{tabular}{lcc}
\toprule
\textbf{Approach} & \textbf{Runtime (s)} & \textbf{Makespan (s)} \\ 
\midrule
Incremental      & 0.047 & 136.28 \\
Time-schedule    & 1.93  & 288.43 \\
\bottomrule
\end{tabular}
\caption{Comparison of runtime and makespan between the Incremental algorithm and the MILP-based Time-schedule approach.}
\label{tab:comparison}
\end{table}

\subsection{Nonlinear MPC}
We assume each vehicle has a wheelbase of $2.5 \mathrm{m}$. 
The steering angle is limited to the range of $-40^\circ$ to $40^\circ$ to reflect realistic steering capability. We also impose saturation limits on the longitudinal acceleration and the steering rate to ensure smooth and physically feasible control inputs, see below. 
\begin{equation}
    \begin{bmatrix} -4 \\ -0.4 \end{bmatrix} \le \begin{bmatrix} a_k \\ \dot\delta_k \end{bmatrix} \le \begin{bmatrix} 3 \\ 0.4 \end{bmatrix}
\end{equation}
These constraints are integrated into the  NMPC formulation to generate safe and dynamically valid trajectories.

The desired velocity is set to $V_d = 3\, \si{\meter\per\second}$. The MPC control horizon is set to 20 steps, and the weight values used in the cost function are summarized in Table \ref{tab:mpc_cost_weights}. 
For this scenario, the conflict region radius is set to $10\,\si{\meter}$, corresponding to a conservative minimum value suitable for high-density traffic conditions. This choice is based on the typical length of urban vehicles ($3$-$6\,\si{\meter}$) and includes two safety margins applied with respect to the nominal maximum vehicle length: a minimum $33\%$ geometric buffer and an additional $33\%$ margin to account for controller inaccuracies, vehicle dynamics, and deviations from ideal lane-following behavior within the intersection \cite{10115056, zhao2023microscopic}.

\begin{figure}
    \centering
    \includegraphics[width=1\linewidth]{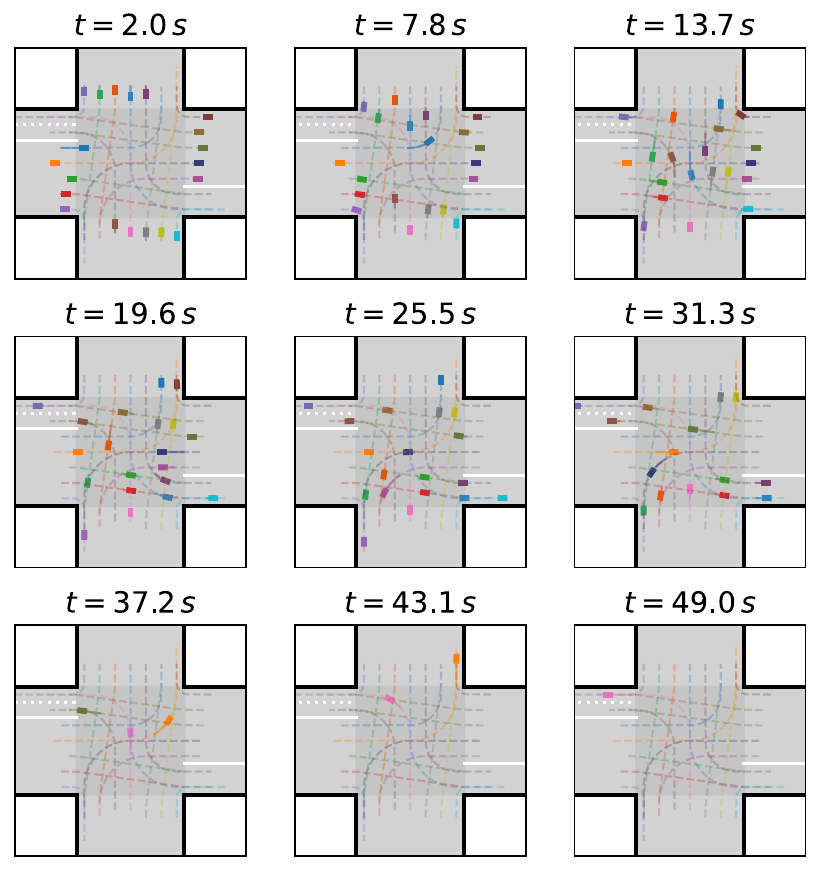}
    \caption{Snapshots in time of the intersection when using the incremental algorithm.}
    \label{fig:snapshots}
\end{figure}

\begin{figure}
    \centering
    \includegraphics[width=1\linewidth]{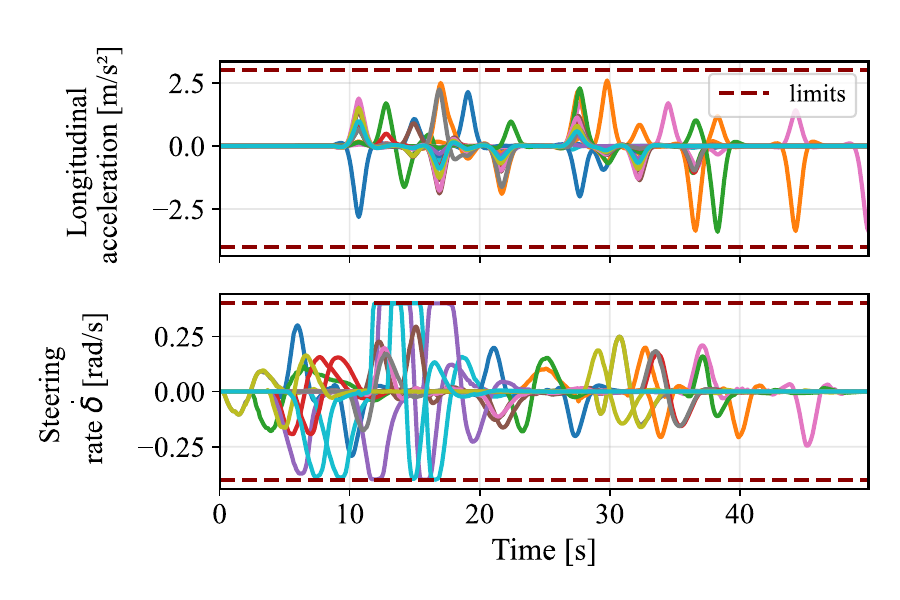}
    \caption{Control inputs for all the vehicles over time. }
    \label{fig:inputs}
\end{figure}

\begin{figure}
    \centering
    \includegraphics[width=\linewidth]{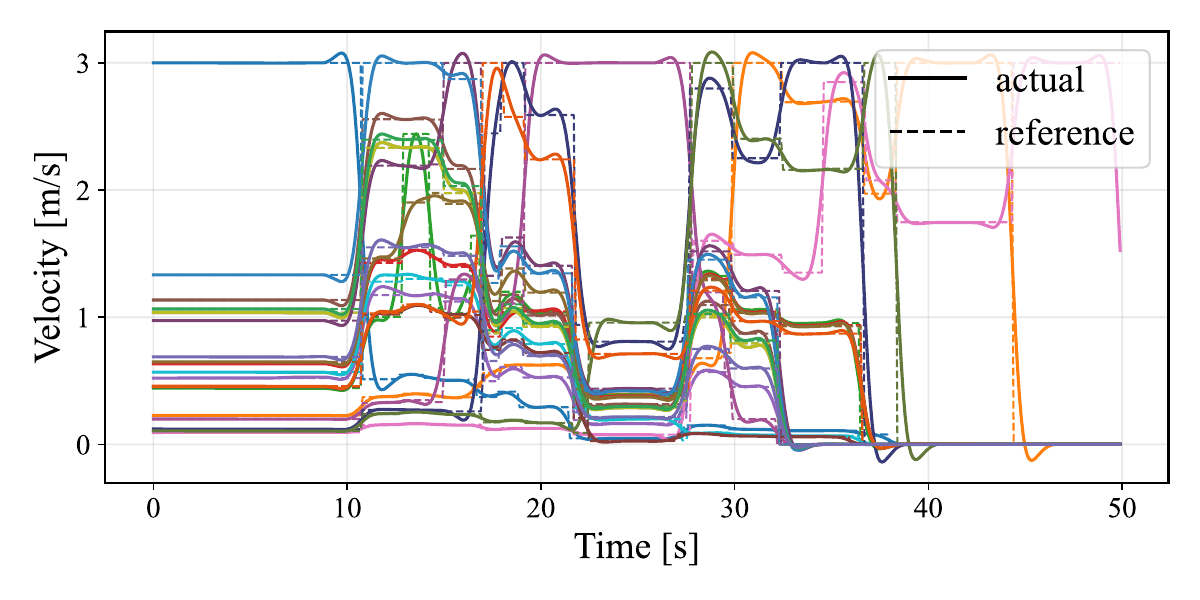}
    \caption{Velocities for all the vehicles over time. }
    \label{fig:vels}
\end{figure}

For the scenario illustrated in Figure~\ref{fig:rw_scen}, the positions of the twenty vehicles are shown during simulation at snapshots in time in Figure~\ref{fig:snapshots}. As observed from the results, all vehicles successfully follow their designated trajectory on their pre-known path and travel through the intersection without any collisions. Additionally, as illustrated in Figure~\ref{fig:inputs}, the control inputs remained within their specified constraints. Furthermore, Figure~\ref{fig:vels} demonstrates that the vehicles closely tracked their reference velocities provided by the planner. As we can see, at each time step, at least one vehicle reaches the maximum speed of 
$V_d = 3\, \si{\meter\per\second}$ in this scenario.
The continuous animation of the simulation results is \href{https://amirrezaaa.github.io/Multi-vehicle-Coordination-at-Intelligent-Intersections-/}{available online}.

\begin{table}
  \begin{tabularx}{\linewidth}{@{}l >{\centering\arraybackslash}X@{}}
    \toprule
    \textbf{Description} & \textbf{Value} \\
    \midrule
    Position tracking cost                 & 50.0 \\
    Velocity tracking cost                 & 10.0 \\
    Heading tracking cost                  & 10.0 \\
    Control effort (acceleration/steering) & 0.5 \\
    Control smoothness (acceleration/steering) & 10.0 \\
    Steering smoothness   & 1.0 \\
    \midrule
    Terminal cost for final position error & 200.0 \\
    Terminal cost for final velocity error & 30.0 \\
    Terminal cost for final heading error  & 30.0 \\
    \bottomrule
  \end{tabularx}
  \caption{Cost function weights used in the NMPC formulation.}
  \label{tab:mpc_cost_weights}
\end{table}

\subsection{Implementation Details}~\label{sec:imp:det}
All simulations and experiments were implemented using Python 3.12 with standard scientific computing libraries. The code was executed on a computer equipped with an Apple M4 chip and 24 GB  RAM. The implementation is entirely in Python and no compilation, parallelization or hardware-acceleration is utilized; the implementation is not optimized for speed. With a compiled or parallelized implementation, notably faster execution times are to be expected.

\section{Conclusions and Future Work}\label{sec6}
In this work, we proposed a novel method for computationally efficient, safe, and near-optimal multi-vehicle trajectory planning in complex traffic scenarios, with a particular focus on intersections. The main contribution of the paper is a planning framework that decomposes a shortest path problem, corresponding to a minimum time problem, in a high-dimensional configuration space into a sequence of 2D graph search problems. We introduced and compared two decomposition strategies, the Incremental approach (Algorithm \ref{alg1}) and the Pairwise approach (Algorithm \ref{alg2}) for safe collision-free trajectory planning. We showed that the computational complexity for both methods is $\mathcal{O}(Nr^2)$, where $N$ is the number of vehicles and $r$ is the number of conflict points, close to which no more than one vehicle can be at the same time.  

To validate the usability of the planned trajectories, we subsequently used NMPC for trajectory tracking. By using the NMPC it was ensured that the planned trajectories can be smoothly and safely followed by vehicles with realistic dynamics. Furthermore we showed that the proposed approach significantly outperforms classic MILP-based time-scheduling, both in terms of runtime and minimum transportation time, i.e., the time it takes under velocity constraints for all vehicles to pass the intersection.  

Our proposed approach assumes predefined reference paths and currently neglects localization errors and the presence of human-driven vehicles in mixed traffic scenarios. This will require accounting for the unpredictability and varying behaviors of human drivers. Future work will address these limitations by exploring extensions of the framework for handling dynamic conflict zones via online re-planning and handling of dynamic conflicts directly at the control layer. 

\bibliographystyle{IEEEtran}
\bibliography{references}

\end{document}